\documentclass[prd,aps,floats,floatfix,superscriptaddress,preprintnumbers,
showpacs,eqsecnum,nofootinbib,twocolumn]{revtex4}
\usepackage{amsfonts,amsmath,amssymb,latexsym,array,
theorem,mathrsfs,subfigure,bm,float,color,graphicx}
\usepackage{multirow}

\definecolor{red}{rgb}{1,0,0}
\definecolor{blue}{rgb}{0,0,1}
\definecolor{green}{rgb}{0,1,0}

\DeclareMathAlphabet{\mathpzc}{OT1}{pzc}{m}{it}

\newcommand{\ma}[1]{\mbox{$\mathcal{#1}$}}

\newcommand{\calhR}[1]{\raisebox{2ex}{\tiny ({\em h})}\hspace{-0.8em}{\ma R}}

\newcommand{\mlv}{m_{\mathrm{LV}}}

\newcommand{\N}{{\rm I\kern-.22em N}}
\newcommand{\Z}{{\sf Z\kern-.42em Z}}
\newcommand{\R}{{\rm I\kern-.22em R}}
\newcommand{\C}{{\rm\kern.22em\rule[.1ex]{.06em}{1.4ex}\kern-.28em C}}
\newcommand{\Q}{{\rm\kern.22em\rule[.1ex]{.06em}{1.4ex}\kern-.28em Q}}
\newcommand{\F}{{\sf F\kern-.42em F}}
\newcommand{\Y}{{\sf Y\kern-.62em Y}}
\newcommand{\X}{{\sf X\kern-.61em X}}
\newcommand{\G}{{\sf G\kern-.61em G}}

\newcommand{\ha}[1]{\mbox{$\hat{#1}$}}
\newcommand{\ch}[1]{\mbox{$\check{#1}$}}

\begin{document}

%<<<<<<<<<<<<< TITLE >>>>>>>>>>>>>>>%
\title{
Stability of Singularity-free Cosmological Solutions in Ho\v{r}ava-Lifshitz Gravity
}

%<<<<<<<<<<<<< AUTHOR >>>>>>>>>>>>>>>%

\author{Yosuke {\sc Misonoh}}
\email{misonoh"at"aoni.waseda.jp}
\address{Department of Physics, Waseda University, 
Okubo 3-4-1, Shinjuku, Tokyo 169-8555, Japan}

\author{Mitsuhiro {\sc Fukushima}}
\email{dark-matter"at"gravity.phys.waseda.ac.jp}
\address{Department of Physics, Waseda University, 
Okubo 3-4-1, Shinjuku, Tokyo 169-8555, Japan}

\author{Shoichiro {\sc Miyashita}}
\email{miyashita"at"gravity.phys.waseda.ac.jp}
\address{Department of Physics, Waseda University, 
Okubo 3-4-1, Shinjuku, Tokyo 169-8555, Japan}

%<<<<<<<<<<<<< ADDRESS >>>>>>>>>>>>>>>%

%<<<<<<<<<<<<< DATE >>>>>>>>>>>>>>>%
\date{\today}

%======================================%
%<<<<<<<<<<<<< ABSTRACT >>>>>>>>>>>>>>>%
%======================================%
%%%%%%%%%%%%%%%%%%%%%%%%%%%%%%%%%%%%%%%%%%%
%%%%%%%%%%%%%%%%%%%%%%%%%%%%%%%%%%%%%%%%%%%
%%%%%%%%%%%%%%%%%%%%%%%%%%%%%%%%%%%%%%%%%%%
%%%%%%%%%%%%%%%%%%%%%%%%%%%%%%%%%%%%%%%%%%%
\begin{abstract}
We study stability of singularity-free cosmological solutions with positive cosmological constant 
based on projectable Ho\v{r}ava-Lifshitz (HL) theory.
In HL theory, the isotropic and homogeneous cosmological solutions with bounce 
can be realized if spacial curvature is non-zero.
By performing perturbation analysis around non-flat Friedmann-Lema\^{i}tre-Robertson-Walker (FLRW) spacetime,
we derive a quadratic action and discuss the stability, i.e, ghost and tachyon-free conditions.
Although the squared effective mass of scalar perturbation must be negative in infrared regime, 
we can avoid tachyon instability by considering strong Hubble friction.
Additionally, we estimate the backreaction from the perturbations on background geometry,
especially, against anisotropic perturbation in closed FLRW spacetime.
It turns out that certain types of bouncing solution may be spoiled even if all perturbation modes are stable. 
\end{abstract}
%%%%%%%%%%%%%%%%%%%%%%%%%%%%%%%%%%%%%%%%%%%
%%%%%%%%%%%%%%%%%%%%%%%%%%%%%%%%%%%%%%%%%%%
%%%%%%%%%%%%%%%%%%%%%%%%%%%%%%%%%%%%%%%%%%%
%%%%%%%%%%%%%%%%%%%%%%%%%%%%%%%%%%%%%%%%%%%

%<<<<<<<<<<<<< PACS NUMBER >>>>>>>>>>>>>>>%

\pacs{04.60.-m, 98.80.-k, 98.80.Cq} 

\maketitle

%======================================%
%<<<<<<<<<<<< SECTION I  >>>>>>>>>>>>>>%
%======================================%
%%%%%%%%%%%%%%%%%%%%%%%%%%%%%%%%%%%%%%%%%%%
%%%%%%%%%%%%%%%%%%%%%%%%%%%%%%%%%%%%%%%%%%%
%%%%%%%%%%%%%%%%%%%%%%%%%%%%%%%%%%%%%%%%%%%
%%%%%%%%%%%%%%%%%%%%%%%%%%%%%%%%%%%%%%%%%%%
\section{Introduction}
Spacetime singularity at the beginning of Universe is a problem of great importance in standard cosmology.
According to singularity theorem proved by Hawking and Penrose\cite{singularity_theorem}, 
a spacetime singularity must be appeared in finite past.
Since the appearance of singularity means breakdown of classical gravitational theory, 
one may expect that fundamental theory beyond General Relativity (GR), i.e., quantum theory of gravity, 
resolves the problem of infiniteness.
A lot of attempts to resolve the singularity at the beginning of Universe 
have been proposed based on extension of GR\cite{bouncing_cosmologies}, 
e.g., superstring theory\cite{ref_superstring}, loop quantum gravity\cite{ref_LQG},
causal dynamical triangulation\cite{ref_CDT} and gravity with non-local operator\cite{ref_non-local}.
Regardless those efforts, the dynamics of very early Universe is still unclear. 
Because we have not achieved complete theory of quantum gravity, yet.

Recently, a gravitational theory attracts attention as a candidate for quantum gravity, 
which is called Ho\v{r}ava-Lifshitz (HL) theory\cite{HL_original}.
The theory is characterized by Lifshitz scaling\cite{ref_LS} which is an anisotropic scaling of spacetime : 
$t \to b^{-1} t$ and $\vec{x} \to b^{-z} \vec{x}$ with dynamical exponent $z$.
If we set $z=3$, 
all types of ultraviolet divergence via Feynman diagrams can be suppressed in four-dimensional spacetime\cite{ref_renormalize_HL}.
It means that gravitational interaction can be renormalized by adding appropriate counterterms.
Thus, the renormalizable gravitational action is composed of second order time derivatives and
up to sixth order spacial derivatives.

Based on HL theory, ultraviolet spacetime structures have been discussed such as
black hole solutions with universal horizon which is a causal boundary for superluminal propagating modes\cite{HL_BHTS}.
In particular, singularity avoidance is an intriguing subject for study.
We expect that quantum gravitational theory resolves spacetime singularities in classical theory of gravity.
The key point to avoid spacetime singularity is violating null energy condition.
As indicated in \cite{matter_bounce_HL}, in Friedmann-Lema\^{i}tre-Robertson-Walker (FLRW) spacetime 
with non-zero spacial curvature,
higher spacial curvature terms in action avoid evolving into singularity. 
Namely, $z=2$ and $z=3$ terms mimic ``dark radiation" and ``dark stiff matter", respectively.
Since the energy densities of these effective matter components depend on 
coupling constants in the theory,
null energy condition can be violated if the values of coupling constants are arbitrary.
As a result, we can find bouncing universe and oscillating universe as singularity-free
solutions\cite{class_HL_cosmology, previous_1, previous_2}.

Although the cosmological singularity avoidance can be realized 
via higher spacial curvatures as ultraviolet modification of gravity, 
one may consider such solutions show unstable behavior.
More specifically, it may possible that the effective matter components derived from $z>1$ Lifshitz scaling terms 
make spacetime unstable because of violation of the energy condition.
It is reasonable that the spacetime around Planck scale is perturbed by quantum fluctuation of gravity. 
Thus, to examine the spacetime stabilities of these singularity-free solutions is indispensable 
in order to construct a cosmological scenario without initial singularity.

In the paper \cite{previous_2}, dynamics of Bianchi type IX spacetime, 
i.e., spacially homogeneous closed cosmological model, is discussed.
In other words, the effect of spacetime anisotropy to singularity-free solutions in closed FLRW spacetime is examined.
From the result, it is found that the stability against small anisotropic perturbation depends on 
the coupling constants of the theory.
Thus, we expect that the dynamics of the other types of perturbations 
are also affected by the coupling constants. 
Since HL theory is renormalizable, 
the values of coupling constants at Planck scale can be evaluated via beta functions from renormalization group, in principle.
If we obtain the values of coupling constants in ultraviolet regime, 
we may predict the dynamics of our Universe at very early stage, which cannot directly observed.
Thus, we set the goal of this paper to show the stability conditions for singularity-free cosmological solutions 
against linear order perturbation not only anisotropic modes but also inhomogeneity ones.

The perturbation analysis in FLRW spacetime in the context of HL theory had been discussed.
Without spacial curvature, there is quite a lot of study, 
e.g., primordial perturbation\cite{HL_pert}, 
stability of scalar perturbation\cite{IR_stability,healty_ex}
 {and stability of de Sitter spacetime}\cite{dS_stability}.
Turning our attention to the case including non-zero spacial curvature, 
the analyses of scalar perturbation have been performed.
In \cite{NF_pert_1}, the authors show scalar perturbation in vacuum FLRW spacetime,
and  in \cite{NF_pert_2}, dynamics of scalar field in bouncing universe is discussed.
Recalling our motivation to investigate spacetime stabilities of singularity-free solutions, 
it is necessary to see dynamics of tensor and vector degree of freedoms as well as scalar ones.

Thus, in this paper, we perform perturbation analysis regarding tensor, vector and scalar modes 
based on projectable HL theory.
Although non-projectable HL theory which is infrared completion of projectable HL theory has been proposed\cite{healty_ex}, 
its renormalizablity is still unclear.
Therefore we focus only on the projectable case because of the renormalizable characteristic.

The rest of this paper is organized as follows : 
In Section \ref{FLRW_HL}, we briefly review projectable HL theory,
especially in FLRW background with non-zero spacial curvature.
The perturbation theory around non-flat FLRW spacetime in HL theory is discussed in
Section \ref{pert_FLRW_HL}.
In Section \ref{stability_FLRW_HL}, we discuss stabilities of bouncing solution
of HL theory in non-flat FLRW spacetime by investigating ghost and tachyon-free conditions.
Additionally, we estimate the backreaction from the perturbation on background geometry,
especially, against anisotropic perturbation in closed FLRW spacetime.
Section \ref{conclusion} is devoted to conclusion of this paper.

%======================================%
%<<<<<<<<<<<< SECTION II  >>>>>>>>>>>>>>%
%======================================%
%%%%%%%%%%%%%%%%%%%%%%%%%%%%%%%%%%%%%%%%%%%
%%%%%%%%%%%%%%%%%%%%%%%%%%%%%%%%%%%%%%%%%%%
%%%%%%%%%%%%%%%%%%%%%%%%%%%%%%%%%%%%%%%%%%%
%%%%%%%%%%%%%%%%%%%%%%%%%%%%%%%%%%%%%%%%%%%

\section{ Ho\v{r}ava-Lifshitz theory in FLRW spacetime}
\label{FLRW_HL}
We briefly review the projectable HL theory, especially, in FLRW spacetime.
The gravitational action is given by\cite{SVW}
 \begin{eqnarray}
  S = {\mlv^2 \over 2} \int dt \ d^3x \, \, \left( \mathcal{L}_K+\mathcal{L}_{P} \right) \,, \label{HL_action}
 \end{eqnarray}
 with
   \begin{eqnarray}
   \mathcal{L}_{K} &:=& N\sqrt{g}  \left( \mathcal{K}_{ij} \mathcal{K}^{ij} -\lambda \mathcal{K}^2  \right)  
 \,,
   \\  \notag \\
 \mathcal{L}_{P}&:=& -N \sqrt{g} \Big[ 
 \mathcal{V}_{z=1} + \mlv^{-2} \mathcal{V}_{z=2} + \mlv^{-4} \mathcal{V}_{z=3}
 \Big]
 \,. \label{potential}
 \end{eqnarray}
 where,  $\mlv$ is Lorentz violating mass scale which may be expected Planck mass.
 The extrinsic curvature $\mathcal{K}_{ij}$ is defined in terms of the lapse function $N$, the shift vector $N_i$ and the three-dimensional 
 induced metric $g_{ij}$ :
 \begin{eqnarray}
 \mathcal{K}_{ij} :=  {1 \over 2N} \left[ \partial_t g_{ij} -\nabla_i  N_j -\nabla_j  N_i  \right]\,.
\end{eqnarray}
where, $\nabla_i$ represents the three-dimensional covariant derivative.
The potential terms are defined by
\begin{eqnarray}
\mathcal{V}_{z=1} &:=& 2\Lambda   + g_1\mathcal{R} \notag \\
\mathcal{V}_{z=2} &:=& g_2 \mathcal{R}^2  + g_3 \mathcal{R}^{i}_{~j} \mathcal{R}^{j}_{~i} \notag \\
\mathcal{V}_{z=3} &:=& g_4 \mathcal{R}^3 + g_5 \mathcal{R}\, \mathcal{R}^{i}_{~j} \mathcal{R}^{j}_{~i} 
+g_6 \mathcal{R}^{i}_{~j} \mathcal{R}^{j}_{~k} \mathcal{R}^{k}_{~i} \notag \\
&&+g_7 \mathcal {R} \nabla^2 \mathcal{R} +{g_8 \nabla_{i} \mathcal{R}_{jk} \nabla^{i} \mathcal{R}^{jk}} \,, \label{potential_p}
\end{eqnarray}
where, $\Lambda$ is the cosmological constant, $\lambda$ and $g_n$ $(n=1$-$8$) are dimensionless coupling constants.
 The potential terms include the higher order spacial curvatures $\mathcal{R}_{ij}$, $\mathcal{R}:=\mathcal{R}^i_{~i}$ 
 up to sixth order spacial derivatives.
In what follows, we adopt the unit $\mlv =1$ unless otherwise noted.

We shall focus on the non-flat FLRW spacetime whose induced metric $g_{ij}$ is given by
\begin{eqnarray}
ds^2 =  a^2 \left[ d \chi ^2 + f(\chi)^2 \, ( d \theta^2 + \sin^2 \theta \, d \phi^2 ) \right] \,.
\end{eqnarray}
with 
\begin{eqnarray}
f(\chi) := 
\begin{cases}
\sin \chi & \mathrm{for}~K=1 \\
\sinh \chi & \mathrm{for}~K=-1 
\end{cases}\,, \label{def_f}
\end{eqnarray}
where, $a$ is a scale factor which depends only on time.
The metrics whose spacial curvature $K=1$ and $-1$ correspond to closed and open FLRW spacetime, respectively.
For closed case, the domains of the variables are defined by $0 \leq \chi  \ {<} \  \pi$,  $0 \leq \theta \ {< } \  \pi$ and 
$0 \leq \phi \ {< } \ 2\pi$.
For open case,  $0 \leq \chi < \infty$, the domains of $\theta$ and $\phi$ are as same as closed ones.

Assuming above ansatz and taking variation with respect to $a$, $N$ and $N_i$, 
we obtain the dynamical equation of scale factor, the Hamiltonian constraint and the momentum constraint, respectively.
Since the momentum constraint gives a trivial relation, we omit it.
The dynamical equation of the scale factor is given by
\begin{eqnarray}
&&{3\lambda -1 \over 2} \left( 2 \dot{H} +  3 H^2 \right) \notag \\
&&~~~~~~~~~~- \left( \Lambda + g_1{K \over a^2} - {g_r \over 3a^4} - {g_s \over a^6}  \right) =0\,,
\label{bg_eom_a}
\end{eqnarray}
with
\begin{eqnarray}
g_r &:=& 6K^2 (3g_2 +g_3)\,, \\ 
g_s &:=& 12K^3 (9g_4 +3g_5 +g_6) \,.
\end{eqnarray}
After taking variation, we have imposed the gauge condition so that $N=1$ and $N_i=0$.
It is notable that the terms derived from the higher spacial curvatures behave as virtual matter fields.
More specifically, the $g_r$ and $g_s$ terms effectively work  as ``radiation" and ``stiff matter", respectively.
Note that the $g_7$ and $g_8$ terms do not affect to the background solution because of the spacetime symmetry.

Turning our attention to the Friedmann equation which corresponds to the Hamiltonian constraint 
in FLRW spacetime. 
In our case, 
we cannot construct the Friedmann equation via taking variation of the action.
Since we have imposed the projectability condition, i.e., $N(t,x) \to N(t)$, 
the Hamiltonian constraint takes the following form : 
\begin{eqnarray}
 \int d^3x \,\left( \mathcal{L}_K -\mathcal{L}_P \right) =0 \,,  \label{global_H_const}
\end{eqnarray}
namely, the Hamiltonian constraint turns to be a global condition instead of local one.
It means that we have to know the information within the entire spacetime 
to construct the Hamiltonian constraint,
and thus, the equation (\ref{global_H_const}) is not viable without special assumption.

Then, we derive the Friedmann-like equation by considering the structure of the basic equations
in FLRW spacetime.
Note that the Friedmann equation (and matter conservation law) basically generates 
the dynamical equation of the scale factor by differentiating with respect to time.
Thus, we can obtain the following equation by performing the time integration of (\ref{bg_eom_a}) :
\begin{eqnarray}
 H^2 - {2 \over 3(3\lambda-1)} \left[ \Lambda + 3g_1{K \over a^2} +{g_r \over a^4} + {g_s \over a^6} \right] = {\mathcal{C} \over a^3}\,. \notag \\ \label{Friedman_eq}
\end{eqnarray}
where, $\mathcal{C}$ is an integration constant.
Note that the $\mathcal{C}$ term behaves as a dust whose energy density is proportional to spacial volume, i.e., $a^{-3}$\cite{DM_integrate}.

For later convenience, we define the following quantities : 
\begin{eqnarray}
\mathcal{E} &:=&  (3\lambda -1) \left( 2 \dot{H} +  3 H^2 \right) \notag \\
&& - 2\left( \Lambda + g_1{K \over a^2} - {g_r \over 3a^4} - {g_s \over a^6}  \right) \,,   \\ \notag \\
\mathcal{H} &:=& { 6(3\lambda-1)} H^2  -  4\left[ \Lambda +3g_1{K \over a^2} +{g_r \over a^4} + {g_s \over a^6} \right] \,. 
\notag \\
\end{eqnarray}
Then, the dynamical equation of the scale factor and Friedmann-like equation can be written by
$\mathcal{E}=0$ and $\mathcal{H} = { 6 (3\lambda -1) } \mathcal{C}/a^3$, respectively.
Since $\mathcal{C}$ does not depend on time, the value of $\mathcal{H}$ is determined by the initial condition.

One may notice that the spacetime dynamics for $\lambda >1/3$ and $\lambda < 1/3$ are completely different, i.e., 
the sign of time derivative terms are flipped.
Since the limit to GR can be obtained by taking $\lambda \to 1$,
we exclude the $\lambda \leq 1/3$ case, in what follows.
Additionally, the value of $g_1$ must be negative to recover the result based on GR
at least at a background level.
Then, we set $g_1 =-1$ in the rest part of this paper by performing a suitable rescaling of time.

%======================================%
%<<<<<<<<<<<< SECTION III  >>>>>>>>>>>>>>%
%======================================%
%%%%%%%%%%%%%%%%%%%%%%%%%%%%%%%%%%%%%%%%%%%
%%%%%%%%%%%%%%%%%%%%%%%%%%%%%%%%%%%%%%%%%%%
%%%%%%%%%%%%%%%%%%%%%%%%%%%%%%%%%%%%%%%%%%%
%%%%%%%%%%%%%%%%%%%%%%%%%%%%%%%%%%%%%%%%%%%

\section{perturbation analysis around non-flat FLRW background}
\label{pert_FLRW_HL}
Since the ADM formalism is employed in this paper, 
we define the perturbed quantities of ADM variables $\delta N$, $\delta N_i$ and $\delta g_{ij}$ as follows : 
\begin{eqnarray}
N &=& \bar{N} + \delta N\,,~ \label{pert_ADM_1} \\
N_i &=& \bar{N}_i + \delta N_i\,,~  \label{pert_ADM_2}  \\
g_{ij} &=& \bar{g}_{ij} + \delta g_{ij} \label{pert_ADM_3} \,,
\end{eqnarray}
where, $\bar{N}$, $ \bar{N}_i$ and $\bar{g}_{ij}$ denote
the background lapse function, shift vector and three-dimensional induced metric, respectively. 
Since we consider the quadratic gravitational action, 
we define $\delta N$, $\delta N_i$ and $\delta g_{ij}$ as follows :
\begin{eqnarray}
\delta N &=& \bar{N} \left[ {\alpha \over \bar{N}} + {1 \over 2} \left({\alpha \over \bar{N}}\right)^2 \right] \,, \notag \\
\delta N_i &=&  \beta_i \,, \notag \\
\delta g_{ij} &=&  h_{ij} +{1 \over 2} \bar{g}^{ab} h_{a i} h_{b j} \,,
\end{eqnarray}
where, $\alpha$, $\beta_i$ and $h_{ij}$ are perturbations of first order.
Furthermore, we define the perturbed quantities with upper indices as 
$h^i_{~j} := \bar{g}^{ia} h_{aj}$, $h^{ij} := \bar{g}^{ia} \bar{g}^{jb} h_{ab}$, $h := \bar{g}^{ab} h_{ab}$ and $\beta^i := \bar{g}^{ia} \beta_a$.
Then, we shall decompose $\alpha$, $\beta_i$ and $h_{ij}$ into
the scalar, vector, tensor modes.

\subsection{spherical and pseudo-spherical harmonics}
In non-flat FLRW background, 
the scalar, vector and tensor perturbations can be decoupled 
by employing spherical or pseudo-spherical harmonics \cite{Lifshitz_Khalatnikov,Ynlm_ref,SandPS}.
More specifically, 
($\chi, \theta, \phi$) dependences of perturbed ADM variables can be expanded by each modes 
of harmonics, which is similar to black hole perturbation theory. 
For example, a scalar function $\Theta$ can be expanded by scalar spherical harmonics $Q^{(n;lm)}$ 
in closed FLRW spacetime :
\begin{eqnarray}
\Theta (t,\chi ,\theta, \phi ) = \sum_{n=1}^{\infty} \sum_{l=0}^{n-1} \sum_{m=-l}^{l} \Theta^{(n;lm)}(t) \, Q^{(n;lm)} (\chi ,\theta, \phi) \,,
\notag \\
\end{eqnarray}
where, $\Theta^{(n;lm)}$ is a coefficient of each ($n;l,m$) modes which depend only on time.
We summarize the definitions of the tensor spherical and pseudo-spherical harmonics 
in Appendix \ref{app_3-harmonics}.

Since we consider the four-dimensional spacetime, 
ten types of independent tensor harmonics must be equipped as a basis set.
In this paper, we employ one of possible orthonormal basis set ${\bf Y}$ as follows : 
\begin{eqnarray}
{{\bf Y}} &=& \Big\{ {Q}\,, {Q}_i\,, {Q}_{ij}\,, {P}_{ij}\,, {S}_{(o)i}\,, {S}_{(e)i}\,, {S}_{(o)ij}\,, {S}_{(e)ij}\,, 
\notag \\
&&~~ {G}_{(o)ij}\,, {G}_{(e)ij}  \Big\} \,,
\end{eqnarray}
When the perturbation in closed (open) FLRW spacetime is considered, 
we refer the quantities with hatted (checked) superscript.
In what follows, we abbreviate these superscripts to unify the discussions\footnote{
For simplicity, we denote $\sum$ to sum up $n$ modes.
To be precisely,  it must be replaced by integration symbol $\int$
when we consider pseudo-spherical ones because $n$ turns to be continuous number.
}.

Turning our attention to the decomposition of ADM variables by harmonics.
The scalar perturbation can be expanded by the scalar harmonics :
\begin{eqnarray}
\alpha^{\mathrm{(scalar)}} &=& \alpha (t)  \label{scalar_N_pert}\,, \\
\beta_i^{\mathrm{(scalar)}} &=& \sum_{n,l,m}  a^2 \left[ \beta^{(n;lm)}_{(Q)} Q^{(n;lm)}_{i} \right] \,, \\
h_{ij}^{\mathrm{(scalar)}} &=& \sum_{n,l,m} a^2 \left[ h^{(n;lm)}_{(Q)} Q^{(n;lm)}_{ij} +h^{(n;lm)}_{(P)} P^{(n;lm)}_{ij}  \right] \,.
\notag \\
\end{eqnarray} 
Note that we do not have to expand the perturbed lapse function in terms of harmonic function. 
Since projectability condition is imposed, the lapse perturbation also depend only on time variable.
The vector perturbation can be expanded by the vector harmonics :
\begin{eqnarray}
\beta_i^{\mathrm{(vector)}} &=& \sum_{n,l,m} a^2 \left[ \beta^{(n;lm)}_{(S;o)} S^{(n;lm)}_{(o) i}+\beta^{(n;lm)}_{(S;e)} S^{(n;lm)}_{(e) i} \right] \,, \notag  \\  \\ 
h_{ij}^{\mathrm{(vector)}} &=& \sum_{n,l,m} a^2 \left[ h^{(n;lm)}_{(S;o)} S^{(n;lm)}_{(o)ij} +h^{(n;lm)}_{(S;e)} S^{(n;lm)}_{(e)ij}  \right] \,. \notag \\ 
\end{eqnarray} 
The lapse function is not perturbed by the vector perturbation.
The tensor perturbation can be expanded by the tensor harmonics :
\begin{eqnarray}
h_{ij}^{\mathrm{(tensor)}} &=& \sum_{n,l,m} a^2 \left[ h^{(n;lm)}_{(G;o)} G^{(n;lm)}_{(o)ij} +h^{(n;lm)}_{(G;e)} G^{(n;lm)}_{(e)ij}  \right] \,. \label{tensor_metric_pert} \notag \\
\end{eqnarray} 
Note that the lapse and shift perturbation do not exist in tensor perturbation. 
In this paper, we eliminate the following perturbations by choosing gauge : 
\begin{eqnarray}
\alpha = h^{(n;lm)}_{(P)}= h^{(n;lm)}_{(S;o)}=h^{(n;lm)}_{(S;e)}= 0\,.
\end{eqnarray}
In Appendix \ref{sec_gauge_fixing}, the gauge structure in FLRW spacetime is summarized.

\subsection{quadratic action}
Turning our attention to the perturbed action at quadratic order by employing (pseudo-)spherical harmonics.
Note that the formulae of harmonic functions we have applied in this part are summarized in Appendix \ref{app_formulae}.

Before performing perturbation analysis,
we shall mention $n =1$ case for closed FLRW.
In this case, the perturbation is given by
\begin{eqnarray}
\beta_i^{(1;00)} = 0\,,~
h_{ij}^{(1;00)} = {1  \over  \sqrt{2}\, \pi} h^{(1;00)}_{(Q)} \bar{g}_{ij}\,.
\end{eqnarray}
One notice that this mode corresponds to just a shift of the scale factor, i.e., $a(t) \to a(t) +\delta a(t)$, 
and thus, we exclude this perturbation mode in what follows.

We firstly consider the perturbation of kinetic terms to clarify the dynamical degree of freedom.
The quadratic kinetic actions for scalar, vector and tensor perturbation are given by 
\begin{eqnarray}
\delta_{(2)} \mathcal{L}_K^{\mathrm{(scalar)}} &=&
a^3 \Bigg[
-{1 \over 2}(3\lambda-1) \dot{h}_{(Q)}^2 \notag \\
&&~~~~
+{3 \over 4}(3\lambda -1) (2\dot{H} + 3H^2  ) h_{(Q)}^2
\notag \\
&&~~~~
-2(3\lambda-1)  {\nu \over \sqrt{3} }  \dot{h}_{(Q)} \beta_{(Q)} \notag \\
&&~~~~
-2\left[ (\lambda-1)\nu^2 +2K  \right] \beta_{(Q)}^2 
\Bigg] \,, \notag \\ \\
\delta_{(2)} \mathcal{L}_K^{\mathrm{(vector)}} &=&
a^3 
(\nu^2 -3K)\beta_{(S)}^2  \,,
 \\
\delta_{(2)} \mathcal{L}_K^{\mathrm{(tensor)}} &=&
  {a^3 \over 2} \dot{h}_{(G)}^2  \,,
\end{eqnarray} 
where, $\nu^2$ is a eigenvalue of the harmonics which is defined 
in terms of the perturbation mode $n \geq 1$ : 
\begin{eqnarray}
&&\nu^2 := 
\begin{cases}
n^2 -1\,,~ n\in \N &\text{for}~ K=1 \\
n^2 +1\,,~ n \in \R &\text{for}~ K=-1 \\
\end{cases}\,. \label{eigan_nu}
\end{eqnarray}
Since the perturbations with different degrees do not mix, we have abbreviated the superscript $(n;lm)$. 
Note that $\beta_{(S)}$ and $h_{(G)}$ are vanished when $l =0$ and $l \leq 1$, respectively. 

Taking variation with respect to $\beta_{(Q)}$ and $\beta_{(S)}$, 
we obtain the following constraint equations.
 \begin{eqnarray}
0 &=& (3\lambda-1){\nu \over \sqrt{3} } \dot{h}_{(Q)} 
+2 \left[ (\lambda-1)\nu^2 + 2K \right]\beta_{(Q)} \,, \notag \\ \\
0&=& 2(\nu^2 -3K) \beta_{(S)}\,, \label{const_beta_S}
\end{eqnarray}
and plugging these relations into the actions, we obtain the simplified quadratic action as follows : 
\begin{eqnarray}
\delta_{(2)} \mathcal{L}_K^{\mathrm{(scalar)}} &=&
a^3 \Bigg[
{(3\lambda-1) (\nu^2 -3K) \over 3 \left[ (\lambda-1)\nu^2  +2K \right] } \dot{h}_{(Q)}^2 
\notag \\ &&
~~~~+{3 \over 4}(3\lambda -1) (2\dot{H} + 3H^2  ) h_{(Q)}^2 
\Bigg] \,,  \notag \\
\end{eqnarray}
and the vector mode is vanished.
Summing up the quadratic potential terms, we finally obtain
\begin{eqnarray}
\delta_{(2)} \mathcal{L}^{\mathrm{(tensor)}} &=& {a^3 \over 2} \Big[ \mathcal{F}_{(G)} \dot{h}_{(G)}^2 - \mathcal{G}_{(G)} h_{(G)}^2 \Big]\,, \label{pert_action_tensor}  \\
\delta_{(2)} \mathcal{L}^{\mathrm{(scalar)}} &=& {a^3 \over 2} \Big[ \mathcal{F}_{(Q)} \dot{h}_{(Q)}^2 - \mathcal{G}_{(Q)} h_{(Q)}^2 \Big]\,, \label{pert_action_scalar}
\end{eqnarray}
where, $\mathcal{F}_{(G)}$ and $\mathcal{G}_{(G)}$ can be regarded as
kinetic term and mass term of tensor perturbation, respectively, which are given by
 \begin{eqnarray}
\mathcal{F}_{(G)} &:=& 1 \,, \\
\notag \\
\mathcal{G}_{(G)} &=& {\nu^2 \over a^2}  +{\nu^2 \over 3a^4} \left[ -{2 g_r \over K} +3g_3 \nu^2  \right] 
\notag \\
&&
+{\nu^2 \over a^6} \bigg[ 
-{g_s \over K} +6 g_{56} K \nu^2 
+g_8 \nu^2 (\nu^2 -2K)
 \bigg]\,. \notag \\ \label{Gg}
\end{eqnarray}  
 where, $g_{56}:=g_5+ g_6$.
 On the other hand, $\mathcal{F}_{(Q)}$ and $\mathcal{G}_{(Q)}$ are regarded as 
 kinetic term and mass term of scalar perturbation :
\begin{eqnarray}
\mathcal{F}_{(Q)} &:=& {2(3\lambda-1) (\nu^2 -3K) \over 3[ (\lambda-1)\nu^2  +2K ]  }\,, \label{Fq}\\
\notag \\
\mathcal{G}_{(Q)} &:=&  -{2 \over 3a^2}(\nu^2 -3K) \notag \\
&&
+{2 \over 27 a^4}(\nu^2 - 3K) \bigg[ {2 g_r \over K^2 }  ( 2\nu^2 - 3K )
+ 3g_3 \nu^2  \bigg] \notag \\
&&+{2 \over 9a^6} (\nu^2 - 3K) \bigg[ {g_s \over K^2} (4\nu^2 -9K) \notag \\
&&~~~~~~~~~
+2 (3g_{56} - {4} g_7)  K\nu^2 \notag \\
&&~~~~~~~~~ 
+ {(-8 g_7 + 3 g_8)} \nu^2 (3\nu^2 - 10K) \notag 
\bigg] \,, \\ \label{Gq}
 \end{eqnarray}  
The terms including Hubble parameter $H$ and its time derivative $\dot{H}$
have been eliminated by applying the background equation of motion $\mathcal{E} =0$.
Note that there does not exist the integration constant $\mathcal{C}$ appeared in 
the Friedmann-like equation $\mathcal{H}= 6(3\lambda-1)\mathcal{C}/a^3$.
It is because the Hamiltonian constant basically arises as a coefficient of the lapse perturbation.
In our case, we have fixed the gauge so that $\alpha =0$,
then, there is no ambiguity in quadratic action.
Additionally, one notice that there are no dynamical scalar modes in the closed FLRW spacetime with $n=2$ 
because $\mathcal{F}_{(Q)}$ is automatically vanished.

Particular attention should be given to the fact the quadratic actions (\ref{pert_action_tensor}) and (\ref{pert_action_scalar}) cannot be reduced into 
the result based on GR even if we take the limit $\lambda \to 1$ and the terms 
from the higher spacial curvature are neglected.
Based on GR, the scalar degree of freedom is absent if we consider the vacuum 
FLRW spacetime.
However, in our case, the scalar perturbations cannot be eliminated when we take such a limit.
This fact is due to the gauge structure shown in Appendix \ref{sec_gauge_fixing}.
More specifically, the time derivatives of scalar perturbation $\dot{h}_{(Q)}$ must be appeared 
because of the Lorentz violation, i.e., the rotation of time direction is not allowed.

In order to clarify the stability conditions for the perturbations, 
we firstly focus on the coefficient of kinetic terms and mass terms.

\subsubsection{ghost avoidance}
We shall concentrate on the terms derived from $\mathcal{L}_K$.
Since $\mathcal{F}_{(G)}$ and $\mathcal{F}_{(Q)}$
are the coefficients of the kinetic terms of each perturbation modes, 
the conditions $\mathcal{F}_{(G)} \geq0$ and $\mathcal{F}_{(Q)} \geq0$
are required to avoid ghost instabilities in tensor and scalar perturbation, respectively.
Since $\mathcal{F}_{(G)} =1$, 
the tensor perturbations do not show ghost instability for any choice of the coupling constants.

On the other hand, the condition for the scalar one is not trivial. 
In closed FLRW spacetime, namely $K=1$, to satisfy $\mathcal{F}_{(Q)} \geq0$ for every $n \geq 2$, we find
\begin{eqnarray}
\lambda \geq 1 \,. \label{closed_stability_lambda}
\end{eqnarray}
One may notice that (\ref{closed_stability_lambda}) is almost same as  
the stability conditions in Minkowski spacetime\cite{previous_1}.
In open FLRW spacetime, namely $K=-1$, 
the stability condition for every $n \geq 1$ mode is given by
\begin{eqnarray}
\lambda  > 2 \,,
\end{eqnarray}
which gives tighter condition than that of closed case.

\subsubsection{tachyon avoidance}
The terms derived from $\mathcal{L}_P$, namely, $\mathcal{G}_{(G)}$ and $\mathcal{G}_{(Q)}$ 
can be regarded as squared masses of tensor and scalar perturbations, respectively.
Therefore, $\mathcal{G}_{(G)} \geq 0$ and $\mathcal{G}_{(Q)} \geq 0$
must be satisfied, otherwise tachyon instability appears.
One can see that equalities $\mathcal{G}_{(G)} =0$ and $\mathcal{G}_{(Q)} = 0$ give
quadratic equations with respect to $a^2$ and $\nu^2$ with the coefficients related with $g_i$.
It means that the ranges for the scale factor in which $\mathcal{G}_{(G)} \geq 0$ and $\mathcal{G}_{(Q)} \geq 0$
can be expressed in terms of the coupling constants, in principle.

In this part, we focus on the infrared stability, 
i.e., the case without effects of higher order spacial curvatures is considered.
Then, the mass terms in the quadratic action are reduced into 
\begin{eqnarray}
\mathcal{G}_{(Q)} \approx -{2 \over 3a^2} (\nu^2 -3K)\,,~
\mathcal{G}_{(G)} \approx {\nu^2 \over a^2} \,. \label{IR_G}
\end{eqnarray}
We see the tensor perturbation mode has positive squared mass, 
while, the scalar one shows opposite sign.
Thus, in infrared regime, the negative squared mass of the scalar perturbations cannot be avoided.
The similar situation has been found in perturbation around Minkowski spacetime\cite{IR_stability,healty_ex}. 
Of course, the negative squared mass does not always lead the tachyon instability.
If the growing time scale for the scalar perturbation is sufficiently small relative to cosmological time scale,
the instability is suppressed.

Another possibility to avoid infrared tachyon instabilities is 
to consider the extended version of HL theory, namely, relaxing projectability condition\cite{healty_ex}.
It is known that additional $(\nabla_i \ln N)^2$ term in extended action 
can stabilize the scalar perturbation at least in flat background.

%======================================%
%<<<<<<<<<<<< SECTION IV  >>>>>>>>>>>>>>%
%======================================%
%%%%%%%%%%%%%%%%%%%%%%%%%%%%%%%%%%%%%%%%%%%
%%%%%%%%%%%%%%%%%%%%%%%%%%%%%%%%%%%%%%%%%%%
%%%%%%%%%%%%%%%%%%%%%%%%%%%%%%%%%%%%%%%%%%%
%%%%%%%%%%%%%%%%%%%%%%%%%%%%%%%%%%%%%%%%%%%
\section{stability analysis of singularity-free solutions}
\label{stability_FLRW_HL}
In this section,  we analyze the stabilities of singularity-free solutions in non-flat FLRW spacetime.
Since the typical scale of the singularity avoidance is expected to be Planck scale,  
we focus only on the solutions which potentially connect to macroscopic universe.
More specifically,  we consider bouncing cosmological solutions with positive cosmological constant, 
which show bouncing behavior at $a=a_T>0$ and turn to  {accelerating} expanding phase.

One may consider the bouncing solution without cosmological constant in open FLRW spacetime 
can also evolve to macroscopic universe with asymptotic Milne expansion.
However, such solutions seem to be unstable because of weak Hubble friction during expanding phase.
As we mentioned in previous, scaler perturbation possesses negative $\mathcal{G}_{G}$ in infrared regime, 
and thus, we exclude the case with non-positive cosmological constant.

\subsection{background solutions}
The classification of the solutions in vacuum FLRW background is performed in the paper\cite{previous_1}.
In that paper, it is found that there are two types of singularity-free solutions.
One is {\it bouncing universe}, that is, the initial contracting universe turns into the expanding phase 
at $a=a_T$, and the universe keeps expansion without finite upper bound of the scale factor.
The other is {\it oscillating universe} whose scale factor is bounded in the range of $0 < a_{\mathrm{min}} \leq a \leq a_{\mathrm{max}} < \infty$.
Then, the universe shows periodic oscillatory behavior without singularity.

The dynamics of the background spacetime can be examined 
via rewritten Friedmann-like equation : 
\begin{eqnarray}
{1 \over 2} \dot{a}^2 + \mathcal{U}(a) =0 \,, \label{FLRW_potential}
\end{eqnarray}
with
\begin{eqnarray}
\mathcal{U}(a) = {1 \over 3\lambda-1} \left[ K - {\Lambda \over 3}{a}^2 -{ g_r \over 3 a^2}  -{ g_s \over 3 a^4}   \right]\,,
\label{FLRW_potential_form}
\end{eqnarray}
For simplicity, we have taken the integration constant $\mathcal{C}=0$.
Since the first term of the left hand side of (\ref{FLRW_potential}) is not negative, 
the possible ranges of the scale factor are where $\mathcal{U} \leq 0$.
Note that $g_r$ and $g_s$ are related with the coupling constants in $\mathcal{L}_P$, 
and then, these values can take both plus and minus sign if the coupling constants are arbitrary.
Therefore, we can consider the situation in which the scale factor is bounded below by
some non-zero minimum value.
It means the universe is forbade to fall down into the singularity. 
We would like to stress that the singularity avoidance is induced 
because the energy condition is effectively violated.

To analyze the background dynamics with $\Lambda > 0$, 
it is convenient to rewrite the potential $\mathcal{U}$ by rescaled variables with respect to $\ell := \sqrt{3/\Lambda}$:
\begin{eqnarray}
\tilde{\mathcal{U}}(\tilde{a}) = {1 \over 3\lambda-1} \left[ K - \tilde{a}^2 -{ \tilde{g}_r \over 3 \tilde{a}^2}  -{ \tilde{g}_s \over 3 \tilde{a}^4}   \right]\,,
\label{U_potential}
\end{eqnarray}
where, $\tilde{a} := a/\ell$, $\tilde{g}_r :=g_r/\ell^2$ and $\tilde{g}_s := g_s /\ell^4$.
Since $\tilde{\mathcal{U}}=0$ is essentially a cubic equation of $\tilde{a}^2$, we obtain three analytic solutions as follows :
\begin{eqnarray}
\left(\tilde{a}^{[K]}_I \right)^2 &:=& \frac{1}{6}\left[2K + {4(K^2-\tilde{g}_r) \over \tilde{\xi}^{[K]}_I } + \tilde{\xi}^{[K]}_I \right] \,,  \label{a_sols}
\end{eqnarray}
with, 
\begin{eqnarray}
\tilde{\xi}^{[K]}_I &:=& \mathrm{pv}~ 2^{2/3} (e^{ 2 \pi i / 3 })^I \Bigg[2K^3 -3\tilde{g}_r K - 9\tilde{g}_s \notag \\
&&~~+ 9\sqrt{ \left(\tilde{g}_s-\tilde{g}^{[K](+) }_s \right) \left(\tilde{g}_s-\tilde{g}^{[K](-) }_s \right) } \Bigg]^{1/3}\,, \notag \\
\\
\tilde{g}^{[K](\pm) }_s  &:=&
{1 \over 9} \left[ 2 K^3  -3 \tilde{g}_r K  \pm 2(K^2 - \tilde{g}_r)^{3/2} \right]  \,,
\end{eqnarray}
where, $I=1,2,3$.
Note that the points at which $\tilde{\mathcal{U}}(\tilde{a})=0$ can be found 
when corresponding { $\tilde{a}_I^{[K]}$} takes real and positive values. 

Additionally, the above roots cannot be applied to the special case with $\tilde{g}_s=0$.
In this instance, we obtain two analytic solutions as follows :
\begin{eqnarray}
\left(\tilde{a}^{[K]}_\pm \right)^2 &:=& \frac{1}{2}\left[K \pm \sqrt{K^2-\frac{4\tilde{g}_r}{3}} \right] \,.  \label{a_sols}
\end{eqnarray}

\subsubsection{closed FLRW ($K=1$)}
\begin{figure}[htbp]
\begin{center}
\includegraphics[width=80mm]{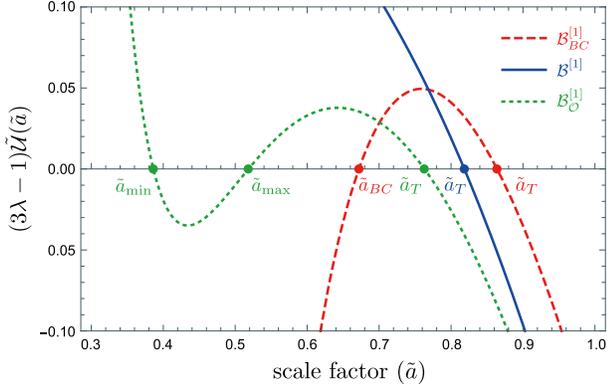}
\end{center}
\caption{
The potentials $\tilde{\mathcal{U}}(\tilde{a})$ for $\mathcal{B}_{BC}^{[1]}$ (red dashed curve), 
$\mathcal{B}^{[1]}$ (blue solid curve) and $\mathcal{B}_{\mathcal{O}}^{[1]}$ (green dotted curve).
The coupling constants are chosen as $\tilde{g}_r=3/10$ and $\tilde{g}_s=2/10$ for $\mathcal{B}_{BC}^{[1]}$, 
$\tilde{g}_r=17/20$ and $\tilde{g}_s=-1/8$ for $\mathcal{B}^{[1]}$, 
$\tilde{g}_r=17/20$ and $\tilde{g}_s=-7/100$ for $\mathcal{B}^{[1]}_{\mathcal{O}}$.
The bouncing radii and the maximum radii of big crunch solution are denoted by $\tilde{a}_T$ and $\tilde{a}_{BC}$, 
respectively.
Additionally, the maximum and minimum radii of oscillation are given by   
$\tilde{a}_{\mathrm{max}}$ and $\tilde{a}_{\mathrm{min}}$.
}
\label{pot_FLRW_K+}
\end{figure}

In closed FLRW universe, namely, $K=1$ case, three types of singularity-free solutions are found. 
We show the typical potentials $\tilde{\mathcal{U}}$ for these solutions in FIG. \ref{pot_FLRW_K+}.
\begin{enumerate}

\item[(a)]{ $\mathcal{B}^{[1]}_{BC}$ } : 
A universe which shows bouncing behavior for initial scale factor $\tilde{a}_{\mathrm{ini}} \geq \tilde{a}_T$, 
however evolves into big crunch for $\tilde{a}_{\mathrm{ini}} \leq \tilde{a}_{BC}$.
We classify this type of the solutions as $\mathcal{B}^{[1]}_{BC}$.
The typical potential is given by the dashed red curve in FIG. \ref{pot_FLRW_K+}.
Note that the domain $\tilde{a}_{BC} < \tilde{a} < \tilde{a}_{T}$ is forbidden.
The solutions $\mathcal{B}^{[1]}_{BC}$ can be found in the following two cases : 
\begin{eqnarray}
 (\mathrm{i})&&0 < \tilde{g}_s < \tilde{g}_s^{[1](+)} \notag \\
 &&~~~~~~~~~~
  \mathrm{with}~\tilde{a}_{BC} = \tilde{a}^{[1]}_2,~\tilde{a}_T = \tilde{a}^{[1]}_3\,. \notag \\ \notag \\
 (\mathrm{ii})&& \tilde{g}_s=0\,,~0<\tilde{g}_r< \frac{3}{4} \notag \\
 &&~~~~~~~~~~
  \mathrm{with}~\tilde{a}_{BC} = \tilde{a}^{[1]}_- ,~\tilde{a}_T = \tilde{a}^{[1]}_+\,. \notag
\end{eqnarray}

\item[(b)]{ $\mathcal{B}^{[1]}$ } : 
A universe which bounce at $\tilde{a}=\tilde{a}_T$
without big-bang singularity for any possible initial scale factor $\tilde{a}_{\mathrm{ini}} \geq \tilde{a}_T$.
We classify this type of the solutions as $\mathcal{B}^{[1]}$.
The typical potential is given by the solid blue curve in FIG. \ref{pot_FLRW_K+}.
The solutions $\mathcal{B}^{[1]}$ can be found in the following three cases : 
\begin{eqnarray}
 (\mathrm{i})&& \tilde{g}_s^{[1](+)}< \tilde{g}_s < 0\,,~\tilde{g}_r<1 ~
  \mathrm{with}~\tilde{a}_T = \tilde{a}^{[1]}_1\,. \notag \\ \notag \\
 (\mathrm{ii})&& 
\begin{cases}
\tilde{g}_s <0 \,,~\tilde{g}_s< \tilde{g}_s^{[1](-)} & \mathrm{for}~~|2\tilde{g}_r-1|<1\\
\tilde{g}_s <0 & \mathrm{for}~~|2\tilde{g}_r-1| \geq 1
\end{cases} \notag \\
 &&~~~~~~~~~~~~~~~~~~~~~~~~~~~~~~~~~~
  \mathrm{with}~\tilde{a}_T = \tilde{a}^{[1]}_3\,. \notag \\ 
(\mathrm{iii})&& \tilde{g}_s=0\,,~\tilde{g}_r \leq 0 ~
 \mathrm{with}~\tilde{a}_T = \tilde{a}^{[1]}_+\,. \notag 
\end{eqnarray}

\item[(c)]{ $\mathcal{B}^{[1]}_{\mathcal{O}}$ } : 
A universe which shows bouncing behavior for initial scale factor $\tilde{a}_{\mathrm{ini}} \geq \tilde{a}_T$,
on the other hand, oscillates if 
$\tilde{a}_{\mathrm{min}}  \leq \tilde{a}_{\mathrm{ini}} \leq \tilde{a}_{\mathrm{max}}$.
The other choice of the initial scale factor is forbidden.
We classify this type of the solutions as $\mathcal{B}^{[1]}_{\mathcal{O}}$.
The typical potential is given by the dotted green curve in FIG. \ref{pot_FLRW_K+}.
The solutions $\mathcal{B}^{[1]}_{\mathcal{O}}$ can be found 
if the following condition is satisfied : 
\begin{eqnarray}
&&\tilde{g}_s <0\,,~\tilde{g}_s^{[1](-)}< \tilde{g}_s < \tilde{g}_s^{[1](+)} \notag \\
&&~~~~~~~
\mathrm{with}~
\tilde{a}_{\mathrm{min}}=\tilde{a}_1^{[1]}\,,~
\tilde{a}_{\mathrm{max}}=\tilde{a}_2^{[1]}\,,~
\tilde{a}_T = \tilde{a}^{[1]}_3\,. \notag
\end{eqnarray}
\end{enumerate}

\subsubsection{open FLRW ($K=-1$) }
\begin{figure}[htbp]
\begin{center}
\includegraphics[width=80mm]{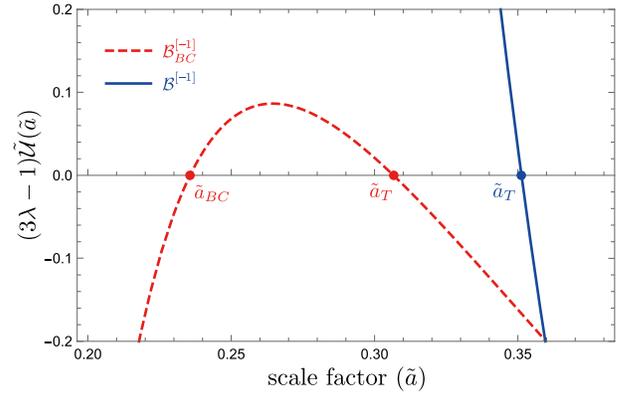}
\end{center}
\caption{
The potentials $\tilde{\mathcal{U}}(\tilde{a})$ for $\mathcal{B}_{BC}^{[-1]}$ (red dashed curve), 
$\mathcal{B}^{[-1]}$ (blue solid curve) in open FLRW universe.
The coupling constants are chosen as $\tilde{g}_r=-1/2$ and $\tilde{g}_s=9/500$ for $\mathcal{B}_{BC}^{[1]}$, 
$\tilde{g}_r=4/5$ and $\tilde{g}_s=-3/20$ for $\mathcal{B}^{[1]}$.
The bouncing radii and the maximum radii of big crunch solution are denoted by $\tilde{a}_T$ and $\tilde{a}_{BC}$, 
respectively.
}
\label{pot_FLRW_K-}
\end{figure}

In open FLRW, namely, $K=-1$ case, two types of singularity-free solutions are found.
We show the typical potentials $\tilde{\mathcal{U}}$ for these solutions in FIG. \ref{pot_FLRW_K-}

 \begin{table*}[tbhp]
\begin{center}
\begin{tabular}{cc|c|c|c|c|c|c|cc}
\hline \hline
\rule[-1.5mm]{0mm}{5mm}& &~$K$~~&~conditions~&~domain~~&~$a_{BC}$&~$a_{\mathrm{mim}}$~&~$a_{\mathrm{max}}$~&~$a_T$&\\
\hline \hline 
\rule[-3mm]{0mm}{8mm}&
\multirow{3}{*}{$\mathcal{B}^{[1]}_{BC}$} & \multirow{3}{*}{$+1$} &~$0 < \tilde{g}_s < \tilde{g}_s^{[1](+)}$~
&~$\tilde{a} \leq \tilde{a}_2^{[1]}$\,,~$\tilde{a} \geq \tilde{a}_3^{[1]}$~
& $\tilde{a}_2^{[1]}$ & N/A  & N/A & ~ $\tilde{a}^{[1]}_3$ \\ \cline{4-10}
 \rule[-3mm]{0mm}{8mm}&& &~$\tilde{g}_s = 0\,,~0<\tilde{g}_r<\frac{3}{4}$~&
~$\tilde{a} \leq \tilde{a}_-^{[1]}$\,,~$\tilde{a} \geq \tilde{a}_+^{[1]}$~&
 $\tilde{a}^{[1]}_-$ &N/A  & N/A & ~$\tilde{a}^{[1]}_+$ \\ \hline
\rule[-3mm]{0mm}{8mm}&
\multirow{7}{*}{$\mathcal{B}^{[1]}$} &\multirow{7}{*}{$+1$} &~$ \tilde{g}_s^{[1](+)}< \tilde{g}_s < 0\,,~\tilde{g}_r<1 $~&
~$\tilde{a} \geq \tilde{a}_1^{[1]}$~&
 N/A  & N/A & N/A & ~$\tilde{a}^{[1]}_1$ \\ \cline{4-10}
\rule[-5mm]{0mm}{12mm}&&&~
$\begin{cases}
  \tilde{g}_s <0 \,,~\tilde{g}_s< \tilde{g}_s^{[1](-)} & \mathrm{for}~~|2\tilde{g}_r-1| < 1\\
 \tilde{g}_s <0 & \mathrm{for}~~|2\tilde{g}_r-1| \geq 1
\end{cases} $~&
~$\tilde{a} \geq \tilde{a}_3^{[1]}$~
& N/A  & N/A & N/A & ~ $\tilde{a}^{[1]}_3$ \\ \cline{4-10}
 \rule[-3mm]{0mm}{8mm}&& &~$\tilde{g}_s = 0\,,~\tilde{g}_r\leq0$~&
~$\tilde{a} \geq \tilde{a}_+^{[1]}$~&
 N/A  & N/A & N/A & ~$\tilde{a}^{[1]}_+$ \\ \hline
\rule[-4mm]{0mm}{10mm}&$\mathcal{B}^{[1]}_{\mathcal{O}}$ & $+1$ &~
$\tilde{g}_s < 0\,,~
\tilde{g}_s^{[1](-)} \leq \tilde{g}_s < \tilde{g}_s^{[1](+)}\,,~0<\tilde{g}_r<1$~&
~$\tilde{a}_1^{[1]}  \leq \tilde{a} \leq  \tilde{a}_2^{[1]}$\,,~$\tilde{a} \geq \tilde{a}_3^{[1]}~$
&N/A& $\tilde{a}^{[1]}_1$ & $\tilde{a}^{[1]}_2$ & ~ $\tilde{a}^{[1]}_3$ \\ \hline
\rule[-3mm]{0mm}{8mm}&$\mathcal{B}^{[-1]}_{BC}$ & $-1$ &~$0 \leq \tilde{g}_s < \tilde{g}_s^{[-1](+)}\,,~\tilde{g}_r<0$~
&~$\tilde{a} \leq \tilde{a}_2^{[-1]}$\,,~$\tilde{a} \geq \tilde{a}_3^{[-1]}$~
&~$\tilde{a}^{[-1]}_2$  & N/A & N/A & ~ $\tilde{a}^{[-1]}_3$ \\ \hline
\rule[-3mm]{0mm}{8mm}&
\multirow{3}{*}{$\mathcal{B}^{[-1]}$} & \multirow{3}{*}{$-1$} &~$\tilde{g}_s < 0$~&
~$\tilde{a} \geq \tilde{a}_3^{[-1]}$~&
 N/A  & N/A & N/A & ~$\tilde{a}^{[-1]}_3$ \\ \cline{4-10}
 \rule[-3mm]{0mm}{8mm}&& &~$\tilde{g}_s = 0\,,~\tilde{g}_r<0 $~&
~$\tilde{a} \geq \tilde{a}_+^{[-1]}$~&
 N/A  & N/A & N/A & ~$\tilde{a}^{[-1]}_+$ \\ 
\hline \hline 
\end{tabular}
\caption{
The conditions and properties of bouncing solutions with positive cosmological constant. 
N/A means that there is no corresponding value of the scale factor.
}
\label{table_solutions}
\end{center}
\end{table*}

\begin{enumerate}
\item[(a)]{ $\mathcal{B}^{[-1]}_{BC}$ } : 
The properties of this solution is quite similar to those of $\mathcal{B}^{[1]}_{BC}$ in closed case.
Namely, this type of the solutions shows bouncing behavior for initial scale factor
$\tilde{a}_{\mathrm{ini}} \geq \tilde{a}_T$, 
on the other hand, evolves into big-bang singularity for $\tilde{a}_{\mathrm{ini}} \leq \tilde{a}_{BC}$.
We classify this type of the solutions as $\mathcal{B}^{[-1]}_{BC}$.
The typical potential is given by the dashed red curve in FIG. \ref{pot_FLRW_K-}.
The solutions $\mathcal{B}^{[-1]}_{BC}$ can be found 
if the following condition is satisfied : 
\begin{eqnarray}
&& 0 < \tilde{g}_s < \tilde{g}_s^{[-1](+)}\,,~ \tilde{g}_r<0 ~\notag \\
&&~~~~~~~~~~~~~~~~~~~~
 \mathrm{with}~
 \tilde{a}_{BC}= \tilde{a}_{2}^{[-1]}\,,~ \tilde{a}_{T}= \tilde{a}_{3}^{[-1]}\,. \notag
\end{eqnarray}

\item[(b)]{ $\mathcal{B}^{[-1]}$ } : 
As is the case with $\mathcal{B}^{[1]}$ in closed FLRW, 
this type of the solutions also shows bouncing behavior for any possible initial scale factor 
$\tilde{a}_{\mathrm{ini}} \geq \tilde{a}_T$.
We classify this type of the solutions as $\mathcal{B}^{[-1]}$.
The typical potential is given by the solid blue curve in FIG. \ref{pot_FLRW_K-}.
The solutions $\mathcal{B}^{[-1]}_{BC}$ can be found in the following two cases : 
\begin{eqnarray}
 (\mathrm{i})&& \tilde{g}_s <0~\mathrm{with}~\tilde{a}_T = \tilde{a}^{[-1]}_3 \,. \notag \\ \notag \\
 (\mathrm{ii})&&\tilde{g}_s =0\,,~ \tilde{g}_r <0 ~\mathrm{with}~\tilde{a}_T = \tilde{a}^{[-1]}_+ \,.  \notag
\end{eqnarray}
\end{enumerate}

\begin{figure}[htbp]
\begin{center}
\includegraphics[width=80mm]{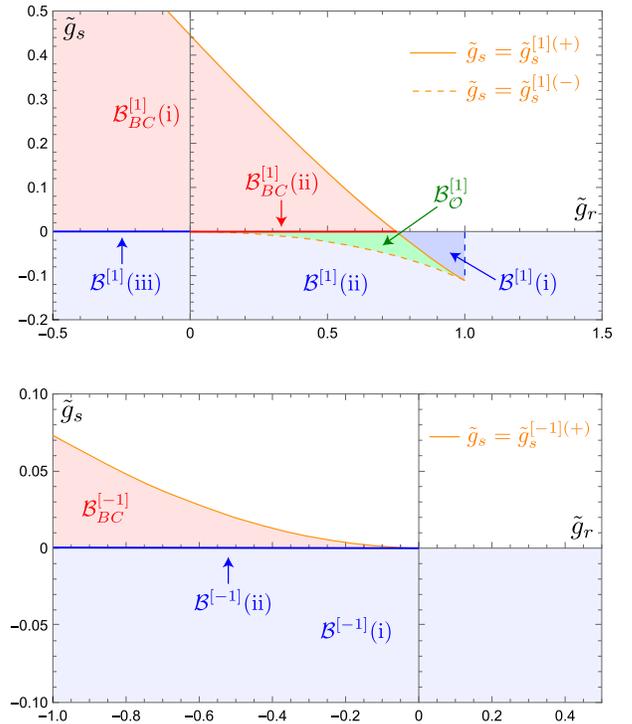}
\end{center}
\caption{
The distribution of the bouncing solutions in $(\tilde{g}_r, \tilde{g}_s)$ plane.
The top and bottom figures correspond the case with $K=1$ and $K=-1$, respectively.
The red, blue and green region indicate the solutions of $\mathcal{B}^{[K]}_{BC}$,
$\mathcal{B}^{[K]}$ and $\mathcal{B}^{[K]}_{\mathcal{O}}$, respectively.
Especially, the solutions with $\tilde{g}_s =0$, 
namely $\mathcal{B}^{[1]}_{BC}$(ii),  $\mathcal{B}^{[1]}$(iii) and $\mathcal{B}^{[-1]}$(ii)
are indicated by colored lines.
}
\label{gr-gs_map}
\end{figure}

We show the properties of singularity-free solutions in TABLE \ref{table_solutions}
and the distribution of the singularity-free solutions in $(\tilde{g}_r, \tilde{g}_s)$ plane 
in FIG. \ref{gr-gs_map}.
\newpage

\subsection{perturbation analysis}
Turning our attention to the ultraviolet stability.
Since the conditions for avoiding ghost instabilities have been discussed in previous section, 
we concentrate on the positivities of $\mathcal{G}_{(Q)}$ and $\mathcal{G}_{(G)}$ 
including higher order spacial curvatures.
Our aim is to clarify the stability conditions for all perturbation modes with finite values of the coupling constants
$\lambda$ and $g_i$ ($i=1$-$8$).
Otherwise the asymptotic safety is violated, 
which may causes the divergence of gravitational force in ultraviolet regime.

\subsubsection{tensor perturbations}
The stability condition for the tensor perturbation is given by $\mathcal{G}_{(G)} \geq 0$.
The positivity of $\mathcal{F}_{(G)}$ is automatically satisfied, namely, 
there is no ghost tensor mode for any choice of coupling constants.
Since $\mathcal{G}_{(G)} = 0$ gives a quadratic equation with respect to $a^2$ and $\nu^2$,
we must require $g_8 \geq 0$.
Otherwise, tensor perturbations with large $\nu^2$ show unstable behavior.

To illustrate the stability of each perturbation mode, 
we firstly analyze the lowest order of the tensor perturbation in closed FLRW background, 
namely, $K=1$ and $n=3$ ($\nu^2=8$) case.
Then, we find
\begin{eqnarray}
\mathcal{G}_{(G)} &=&
{8 \over a^2}  +{16 \over 3 a^4} \left( 12g_3 -g_r \right) 
\notag \\
&&
+{8 \over a^6} \Big[ 
48(g_{56} +g_8) -g_s
 \Big]
\,. \label{tensor_n3}
\end{eqnarray}
In order to stabilize this perturbation mode for any $a>0$, 
the following condition must be satisfied :  
\begin{eqnarray}
\begin{cases}
-{1 \over 9}(g_r -12 g_3)^2 \\ 
~~~~~~~
\geq g_s -48(g_{56}+g_8)\,, & \text{for}~g_r \geq 12 g_3 \,, \\[3mm]
0 \geq g_s -48(g_{56}+g_8)\,, & \text{for}~g_r < 12 g_3 \,. 
\end{cases}  \label{stability_Bianchi_IX_pert}
\end{eqnarray}
One may notice that the stability condition (\ref{stability_Bianchi_IX_pert}) reproduces
the result shown in \cite{previous_2}.
In fact, the tensor perturbations with $n=3$ in closed FLRW background include
homogeneous and anisotropic perturbation, namely, Bianchi type IX spacetime with small anisotropy 
(see Section \ref{anisotropy_pert}).

It is obvious that we should impose the positivity of  $\mathcal{G}_{(G)}$ 
for any perturbation mode $\nu^2 $ to ensure the stability of spacetime against tensor perturbation. 
Of course, we can express the stability conditions for tensor perturbation for any viable $a>0$ and
perturbation mode $\nu^2$ in terms of the coupling constant $g_i$.
Since $\mathcal{G}_{(G)}=0$ gives a quadratic equation in terms of $a^2$ and $\nu^2$, we can solve it, in principle.
However, it is found that the explicit form is quite complicated.
Especially, in closed background, $\nu^2$ takes discrete value and 
this fact complicates the analysis.
Thus, instead of considering general case, 
we show a special case, namely, $g_3=0$ in open FLRW spacetime.
Then,  we find that every mode of tensor perturbation can be stabilized for any $a\geq 0$, 
if one of the following four condition is satisfied : 
\begin{eqnarray}
(\text{i}) && 0 < g_8 \leq g_{56}\,,~g_r >0  \,,~g_s \geq {(g_8-3g_{56})^2 \over g_8} \notag \\
(\text{ii}) && 0 < g_8 \leq g_{56}\,,~g_r \leq 0 \,,~g_s \geq {(g_8-3g_{56})^2 \over g_8} +{g_r^2 \over 9}  \notag \\
(\text{iii}) && g_{56} < g_8\,,~g_r >0 \,,~g_s \geq 4(3g_{56}-2g_{8})  \notag \\
(\text{iv}) && g_{56} < g_8\,,~g_r \leq 0 \,,~g_s \geq 4(3g_{56}-2g_{8})  +{g_r^2 \over 9} \notag 
\end{eqnarray}

Note that there is some difficulty in reconciling above conditions with the conditions for bouncing solutions
in open FLRW spacetime.
Referring TABLE \ref{table_solutions} and FIG. \ref{gr-gs_map}, 
one can see that $\mathcal{B}^{[-1]}_{BC}$ is appeared if ${g_r}<0$ and $g_s>0$ with $|\tilde{g}_r| \gg |\tilde{g}_s|$.
On the other hand, $\mathcal{B}^{[-1]}$ can be found if $g_s<0$.
Then, we shall examine compatibilities of these conditions.

\begin{enumerate}
\item[(i)] The condition (i) is incompatible with both types of bouncing solutions.
Because, $g_r$ and $g_s$ are constrained to be positive, which completely contradicts 
both condition for $\mathcal{B}^{[-1]}_{BC}$ and $\mathcal{B}^{[-1]}$.
\item[(ii)]
Since $g_s$ is constrained to be positive number under the condition (ii), 
there is no stable bouncing solution $\mathcal{B}^{[-1]}$.
Additionally, one may notice that the lower bound of $g_s$ is given in terms of $g_r^2$.
Since $|\tilde{g}_r| \gg |\tilde{g}_s|$ should be satisfied to stabilize the solution $\mathcal{B}^{[-1]}_{BC}$, 
the condition (ii) is hard to compatible unless small $\Lambda >0$ is set.
As we shown later, small $\Lambda >0$ is not preferable to stabilize scalar perturbation
at infrared regime.
\item[(iii)]
Since $g_r$ is constrained to be positive,
the condition (iii) cannot be compatible with the solution $\mathcal{B}^{[-1]}_{BC}$. 
The bouncing solution $\mathcal{B}^{[-1]}$ can be stabilized only if $3g_{56}>2g_{8}$.
\item[(iv)] 
Under the condition (iv), both $\mathcal{B}^{[-1]}_{BC}$ and $\mathcal{B}^{[-1]}$ can be stabilized.
Since $g_s$ is bounded below by $g_r^2$, 
it may require a certain level of tuning to stabilize the bouncing solution $\mathcal{B}^{[-1]}_{BC}$
for the same reason of the condition (ii). 
\end{enumerate}

In general case, we also anticipate unstable tensor perturbations 
in bouncing open FLRW spacetime.
From (\ref{Gg}), one can see that $g_s> 0$ and $g_r > 0$ are preferable 
in order to ensure the positivity of $\mathcal{G}_{(G)}$,
which are basically contradict to the conditions of bouncing solutions 
$\mathcal{B}^{[-1]}_{BC}$ and $\mathcal{B}^{[-1]}$.
Conversely, in closed FLRW spacetime, the tensor perturbations can be
stabilized without special tuning.
Naively, $g_r<0$ and $g_s <0$ are preferred to satisfy the positivity of $\mathcal{G}_{(G)}$
for any $a>0$ and $\nu^2$.
Referring TABLE \ref{table_solutions} and FIG. \ref{gr-gs_map}, 
we find $\mathcal{B}^{[1]}$(ii) solution is under such a condition.

\begin{table*}[t]
\begin{center}
\begin{tabular}{cc|c|c|c|c|c|c|c|c|c|c|c|c|c|cc}
\hline \hline
\rule[-1mm]{0mm}{4mm}&&~type~~&~$\Lambda$~~&~$g_2$~&~$g_3$~~&~$g_4$~&~$g_5$~&~$g_6$~&~~$g_7$~&
~~$g_8$~&~~$g_r$~&~~$g_s$~&~~$a_T$~&~$a_{\mathrm{crit}}$~&~$a_{\mathrm{crit}}/a_{T}$ 
&\\
\hline \hline 
\rule[-3.5mm]{0mm}{9mm}& (i)~~&$\mathcal{B}_{BC}^{[1]}$~&
~~$\displaystyle{{3 \over 2}}$~~&~~$-1$~~&~~$1$~~&~~$\displaystyle{{1 \over 5}}$~~&~~$\displaystyle{-{1 \over 2}}$~~&~~$1$~~&~~$-1$~~&~~$1$~~&~~$-12$~~&~~$\displaystyle{{78 \over 5}}$~~&
~~$1.856$~~&~~$  {3.544}$~~&~~$  {1.909}$~
&
 \\ \hline
\rule[-3.5mm]{0mm}{9mm}& (ii)~~&$\mathcal{B}^{[1]}$~&
~~$\displaystyle{1}$~~&~~$\displaystyle{-{29 \over 90}}$~~&~~$1$~~&$\displaystyle{-{7 \over 108}}$~~&~~$\displaystyle{{1 \over 2}}$~~&~~$-1$~~&~~$-1$~~&~~$1$~~&~~$\displaystyle{{1 \over 5}}$~~&~~$\displaystyle{-1}$~~&
~~$1.744$~~&~~$  {4.699}$~~&~~$  {2.694}$~
&
 \\ \hline
\rule[-3.5mm]{0mm}{9mm}& (iii)~~&$\mathcal{B}^{[1]}_{\mathcal{O}}$~&
~~$\displaystyle{{1 \over 5}}$~~&~~$\displaystyle{{1 \over 5}}$~~&~~$1$~~&~~$0$~~&~~$\displaystyle{-{1 \over 5}}$~~&~~$\displaystyle{{1 \over 4}}$~~&~~$\displaystyle{{1 \over 4}}$~~&~~$1$~~&~~$\displaystyle{{48 \over 5}}$~~&~~$\displaystyle{-{21 \over 5}}$~~&
~~$3.270$~~&~~$  {5.514}$~~&~~$  {1.686}$~
& 
 \\ \hline
\rule[-3.5mm]{0mm}{9mm} & (iv)~~&$\mathcal{B}^{[-1]}_{BC}$~&
~~$\displaystyle{{3 \over 2}}$~~&~~$\displaystyle{-{5 \over 18}}$~~&~~$\displaystyle{{2 \over 3}}$~~&$\displaystyle{-{133 \over 2160}}$~~&~~$\displaystyle{{1 \over 4}}$~~&~~$\displaystyle{-{1 \over 5}}$~~&~~$\displaystyle{{1 \over 16}}$~~&~~$\displaystyle{{1 \over 3}}$~~&~~$-1$~~&~~$\displaystyle{{1 \over 20}}$~~&
~~${0.485}$~~&~~$  {1.380}$~~&~~$  {2.849}$~
&
 \\ \hline
\rule[-3.5mm]{0mm}{9mm}&(v)~~&$\mathcal{B}^{[-1]}$~&
~~$1$~~&~~$\displaystyle{{1 \over 18}}$~~&~~$\displaystyle{-{2 \over 15}}$~~&~~$\displaystyle{{31 \over 108}}$~~&~~$\displaystyle{-{1 \over 2}}$~~&~~$-1$~~&~~$\displaystyle{{1 \over 900}}$~~&~~$\displaystyle{{2 \over 225}}$~~&~~$\displaystyle{{1 \over 5}}$~~&~~$-1$~~&
~~$0.712$~~&~~$  {0.920}$~~&~~$  {1.293}$~
 \\ 
\hline \hline 
\end{tabular}
\caption{
The examples for stable bouncing solutions with positive cosmological constants.
}
\label{table_stable_sol}
\end{center}
\end{table*}
\subsubsection{scalar perturbation}
Although the negative sign of $\mathcal{G}_{(Q)}$ in infrared regime cannot be avoided, 
$\mathcal{G}_{(Q)}$ may take positive value in the deep ultraviolet region 
in which the effects of higher spacial curvatures are predominant.
To realize stable bouncing phase, 
we require the positivity of $\mathcal{G}_{(Q)}$, at least in the range $[0 , a_{\mathrm{ini}}]$
with $a_{\mathrm{ini}} > a_T$.
Since $\mathcal{G}_{(Q)} =0$ gives a quadratic equation in terms of $a^2$ with upward convex, 
the positivity of $\mathcal{G}_{(Q)}$ at least in $[0 , a_{\mathrm{ini}}]$ is guaranteed 
if both of the following conditions are satisfied : 
\begin{eqnarray}
0 &\leq& 
-9a_{\mathrm{ini}}^4
+\bigg[ {2 g_r \over K^2 }  ( 2\nu^2 - 3K )
+ 3g_3 \nu^2  \bigg]a_{\mathrm{ini}}^2 \notag \\
&&+3\bigg[ {g_s \over K^2} (4\nu^2 -9K)
+2 (3g_{56} -  {4} g_7)  K\nu^2 \notag \\
&&~~~~
+ {(-8 g_7 + 3 g_8)} \nu^2 (3\nu^2 - 10K)
\bigg]  \,, \\  \notag \\
0 &\leq& 
{g_s \over K^2} (4\nu^2 -9K)
+2 (3g_{56} -  {4} g_7)  K\nu^2 \notag \\
&&~~~~
+ {(-8 g_7 + 3 g_8)} \nu^2 (3\nu^2 - 10K) 
 \,,
\end{eqnarray}  
Note that the case with $g_8 < 8g_7/3$ must be excluded, 
otherwise the scalar perturbations with large $\nu^2$ shows unstable behavior.  \\

In our analysis, we can find bouncing solutions $\mathcal{B}^{[K]}_{BC}$, $\mathcal{B}^{[K]}$ and 
$\mathcal{B}^{[K]}_{\mathcal{O}}$ which satisfy both tensor and scalar stability conditions,
namely, with $\mathcal{G}_{(G)} \geq 0$ in $[0, \infty)$, and $\mathcal{G}_{(Q)} \geq 0$ at least in $[0, a_{T}]$.
The examples of such solutions are listed in TABLE \ref{table_stable_sol}.
Note that a certain level of tuning is required to find stable open bouncing universe,
because the tensor perturbation tend to be unstable for the reason we mentioned. 

One may wonder about the appropriate value of $a_{\mathrm{ini}}$.
The initial scale factor $a_{\mathrm{ini}}$ seems to relate with
a quantum creation of the universe.
Therefore, it is natural to consider the typical energy scale is estimated at Planck scale.
On the other hand, the typical scale of the bouncing radius is also expected to be around at Planck scale.
Because, the bouncing behavior is induced by higher spacial curvature terms, 
namely, quantum gravitational corrections.
More specifically, we assume the three-Ricci curvature represents 
the energy scale, namely, $m \sim \sqrt{\mathcal{R}} \propto a^{-1}$. 
Then, the ratio of quantum creation scale $m_{\mathrm{ini}}$ to bouncing scale $m_T$ is given by
\begin{eqnarray}
{m_{\mathrm{ini}} \over m_{T} } \sim  
{a_T \over a_{\mathrm{ini}}}  
\,.
\end{eqnarray}
It is natural to consider the ratio is order one.

Additionally, we consider the upper limit of $a_{\mathrm{ini}}$ to satisfy 
the positivity of $\mathcal{G}_{(Q)}$ during bouncing phase.
As we noted, $\mathcal{G}_{(Q)}$ must be negative for large $a$.
Thus, there exists a critical value of the scale factor $a_{\mathrm{crit}}$.
Namely, any scalar perturbation mode possess positive squared mass for $a> a_{\mathrm{crit}}$, 
however, any one of scalar perturbation mode turns to be zero at $a= a_{\mathrm{crit}}$.
Obviously, $a_{\mathrm{ini}}$ must be in $(a_T, a_{\mathrm{crit}})$.
Further constraint for $a_{\mathrm{ini}}$ can be imposed by considering the dynamics of the perturbations.

\subsection{dynamics of perturbations}
To construct a scenario for the non-singular cosmological evolution, 
we have to pay attention to the dynamics of the perturbations.
Taking variation of the quadratic actions with respect to $h_{(G)}$ and $h_{(Q)}$, 
we obtain the  equations of motion for tensor and scalar perturbations, respectively :
\begin{eqnarray}
&&\ddot{h}_{(G)} +3 H \dot{h}_{(G)} + \mathcal{M}_{(G)}^2  {h}_{(G)} =0 \label{G_eom} \,, \\
&&\ddot{h}_{(Q)} +3 H \dot{h}_{(Q)} + \mathcal{M}_{(Q)}^2  {h}_{(Q)} =0 \label{Q_eom} \,,
\end{eqnarray}
where, we define effective squared masses of the tensor and scalar perturbation as
\begin{eqnarray}
\mathcal{M}_{(G)}^2 := {\mathcal{G}_{(G)}  \over \mathcal{F}_{(G)}}\,,~
\mathcal{M}_{(Q)}^2 := {\mathcal{G}_{(Q)}  \over \mathcal{F}_{(Q)}} \,.
\end{eqnarray}

We firstly consider the contracting phase before bounce. 
In this era, the perturbations feel a {\it Hubble acceleration} which is derived from the second terms in 
(\ref{G_eom}) and (\ref{Q_eom}) because of negative Hubble parameter $H<0$.
Since the Hubble acceleration enhances the both perturbation modes, 
this effect should be suppressed by effective mass terms.

Intuitively, the magnitudes of the Hubble accelerations for tensor and scalar perturbation
are given by $H^2$.
Thus, to suppress the unstable behavior, we require 
\begin{eqnarray}
\mathcal{M}_{(Q)}^2 \gtrsim H^2 \,,~\mathcal{M}_{(G)}^2 \gtrsim H^2 \,. \label{contracting_phase_stability}
\end{eqnarray}
throughout contracting phase.
This condition gives a further constraint on possible value of the initial scale factor. 
Namely, $a_{\mathrm{ini}}$ should be in the range of $(a_{T}, a_{H})$, 
where, $a_{H}$ is a value of scale factor in which any one of effective mass of perturbation turns to be $\mathcal{M}^2 = H^2$.
Namely, for $a> a_H$, every perturbation modes shows positive effective squared mass which is larger than $H^2$.
Then, the condition (\ref{contracting_phase_stability}) is ensured in contracting era for $a_{T} < a_{\mathrm{ini}} < a_{H}$.

Note that the possible range for the initial scale factor can be broadened by tuning the value of coupling constant $\lambda$.
Namely, the dependence of $\lambda$ in effective mass-Hubble parameter ratios can be evaluated as follows :
\begin{eqnarray}
{\mathcal{M}_{(G)}^2 \over H^2} \propto 3\lambda-1\,,~
{\mathcal{M}_{(Q)}^2 \over H^2} \propto \lambda-1\,,  \label{mass-Hubble-ratio_lambda}
\end{eqnarray}
then, one can see that large value of $\lambda$ weakens the effect of Hubble acceleration
in both cases.

After the bounce at $a_T$, the universe turns to expand.
As we mentioned,
the effective squared mass of the scalar perturbation must be negative in infrared regime.
Thus, to stabilize the perturbation, 
Hubble friction with $H>0$ must overcome the effects of the negative squared mass of scalar modes.
Namely, we require
\begin{eqnarray}
\left| \mathcal{M}_{(Q)}^2 \right| \lesssim H^2~~\mathrm{when}~~\mathcal{M}^2_{(Q)}<0 \,, \label{expanding_phase_stability}
\end{eqnarray}
in expanding era with large value of the scale factor.

\begin{figure}[htbp]
\begin{center}
\includegraphics[width=80mm]{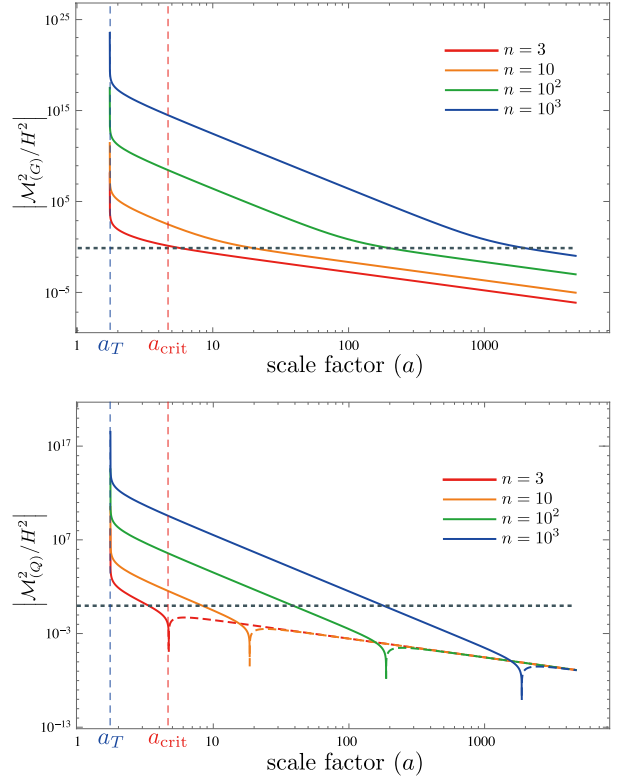} 
\end{center}
\caption{
The typical example of a bouncing solution without any instability after bounce.
In this figure, we show the evolutions of the solution (ii) 
listed in TABLE \ref{table_stable_sol}.
We set the coupling constant $\lambda$ to be unity.
The red, orange, green and blue curves indicate the ratio of the squared effective masses to squared Hubble parameter with
$n=3$, $10$, $10^2$ and $10^3$
($\nu^2 =n^2-1$), respectively.
The top and bottom figure shows those of tensor and scalar perturbations, respectively.
The solid (dashed) curve shows the evolution with $\mathcal{M}^2>0$ ($\mathcal{M}^2<0$).
The gray dotted line indicates $\mathcal{M}^2 = H^2$.
}
\label{B+1_plot}
\end{figure}

We examine the stability of the bouncing solutions by showing concrete examples.
Firstly, we shall show the bouncing solution without any instability throughout the evolution. 
In FIG. \ref{B+1_plot}, we show the evolutions of $\mathcal{M}_{(G)}^2/H^2$ and $\mathcal{M}_{(Q)}^2/H^2$ in terms of scale factor.
From these figures, we find that $\mathcal{M}_{(G)}^2$ of all tensor perturbation modes are always positive, 
and the condition (\ref{expanding_phase_stability}) is always satisfied after turning $\mathcal{M}_{(Q)}^2$ 
to be negative for any scalar perturbation modes.
Thus, the instabilities in scalar perturbation are suppressed by Hubble friction, in this case.

\begin{figure}[htbp]
\begin{center}
\includegraphics[width=80mm]{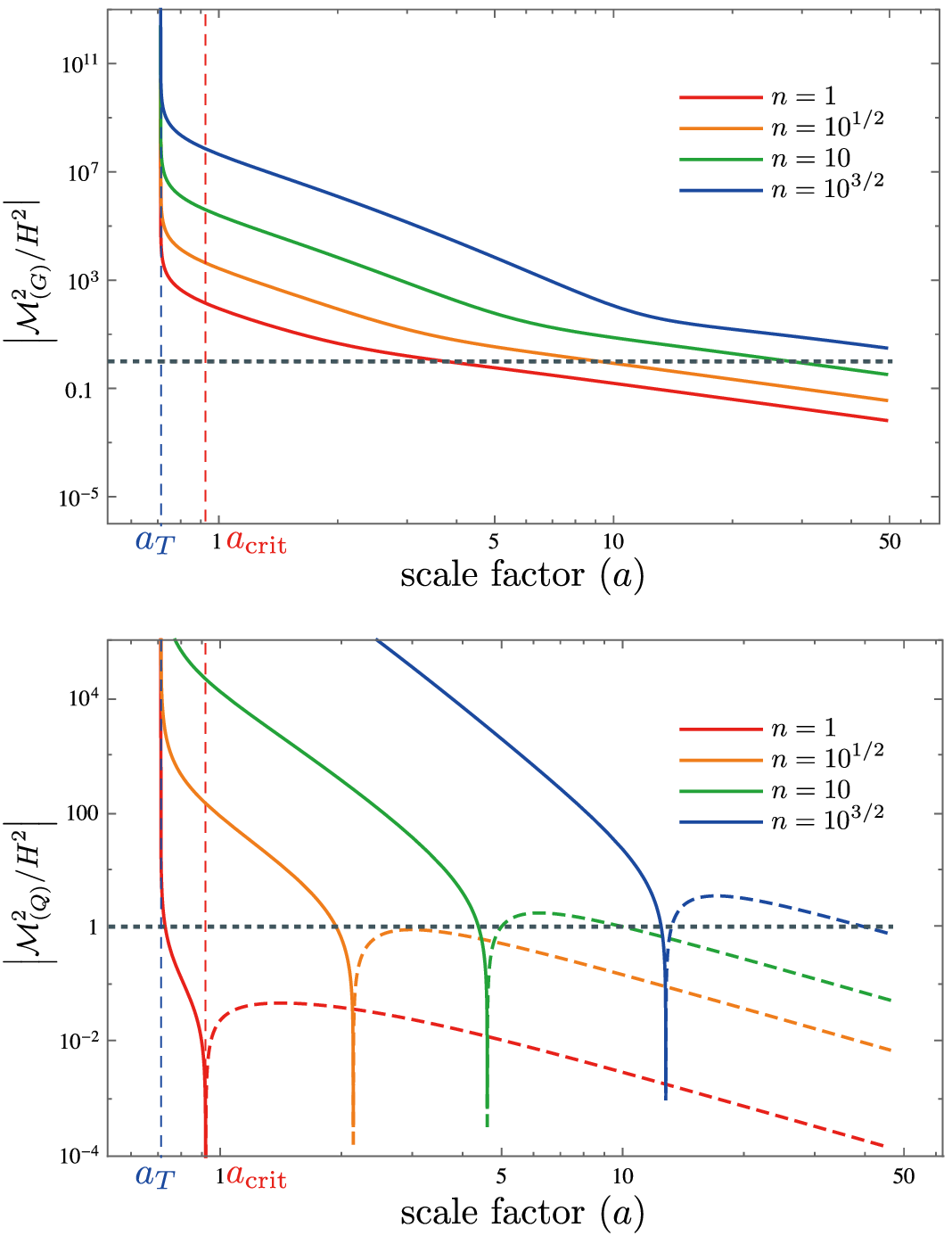}
\end{center}
\caption{
The typical example of a bouncing solution with temporal scalar instabilities.
In this figure, we show the evolutions of the solution (v) listed in TABLE \ref{table_stable_sol}.
We set the coupling constant $\lambda =2.1$.
The red, orange, green and blue curves indicate the ratio of the squared effective masses to squared Hubble parameter with
 $n=1$, $10^{1/2}$, $10$ and $10^{3/2}$ ($\nu^2 =n^2+1$), respectively.
The top and bottom figure shows those of tensor and scalar perturbations, respectively.
The solid (dashed) curve shows the evolution with $\mathcal{M}^2>0$ ($\mathcal{M}^2<0$).
The gray dotted line indicates $\mathcal{M}^2 = H^2$.
}
\label{B-1_plot}
\end{figure}

On the other hand, we show an example of bouncing solution with temporal scalar instabilities in FIG. \ref{B-1_plot}.
Although all of the tensor perturbation shows stable behavior because of positive $\mathcal{M}_{(G)}^2$, 
some of scalar perturbations temporally show tachyon instabilities.
Namely, for a temporary period, 
some  scalar perturbation modes violate the condition (\ref{expanding_phase_stability}).

One may wonder about the growth of scalar perturbation during tachyon instability.
Seeing equation of motion for scalar perturbation, 
it is natural to speculate that the growth rate is related with 
the minimum value of $\mathcal{M}_{(Q)}^2/H^2$.
Thus, we firstly clarify the minimum value of the squared effective mass $\mathcal{M}_{(Q)}^2$.
For simplicity, we consider the asymptotic region, i.e., for large perturbation mode $\nu^2$.
In this limit, the kinetic and mass terms of scalar perturbations given by (\ref{Fq}) and (\ref{Gq}) 
are reduced into the following forms :  
\begin{eqnarray}
\mathcal{F}_{(Q)} &\approx& {2(3\lambda -1) \over 3(\lambda-1) } \,, \\
\mathcal{G}_{(Q)} &\approx&  -{2\nu^2 \over 3a^2} 
+{2 \nu^4 \over 27 a^4}  \bigg[ {4 g_r \over K^2 } + 3g_3 \bigg] 
+{  {2} \nu^6 \over 3a^6}  (3 g_8 -8 g_7 ) \,. \notag \\
 \end{eqnarray}  
 Then, the minimum value of $\mathcal{M}_{(Q)}^2$ is given by
 \begin{eqnarray}
&&\min_{a} \mathcal{M}_{(Q)}^2 \approx
-{3K^2(\lambda -1) \over (3\lambda-1) \eta^2} 
\left[ 2\eta -(4g_r+3g_3 K^2)  \right] \notag \\ \notag \\
&&\text{with}~~
\eta := 4g_r + 3g_3 K^2 \notag \\ 
&&~~~~~~~~~~~~~
+\sqrt{ {243}(3g_8-8g_7)K^4 +(4g_r + 3g_3 K^2)^2}\,. \notag \\ \label{min_value}
\end{eqnarray}
After taking above value, $\mathcal{M}_{(Q)}^2$ monotonically increases with time and approaches to zero.
Then, the scalar perturbation is stabilized by Hubble friction.
The important point is that the minimum value does not depend on the perturbation mode $\nu^2$ in this limit.
Thus, we can conclude that the minimum value of $\mathcal{M}_{(Q)}^2$ can be bounded in finite value.
Namely, if the ratio to Hubble parameter for large $\nu^2$,
\begin{eqnarray}
&&\min_a {\mathcal{M}_{(Q)}^2 \over H^2}  \approx
-{9 K^2 (\lambda -1) \over 2\Lambda \eta^2}
 \left[ 2\eta -(4g_r+3g_3 K^2) \right]\,, \notag \\
\label{MH_rate}
\end{eqnarray} 
is sufficiently suppressed, it is expected that 
there is no serious instability at least at the classical level.
Note that the large value of the positive cosmological constant $\Lambda$  and/or the small value of $\lambda$ decreases above value.

In above discussion, we limited our analysis to the case with large $\nu^2$.
However, we can investigate the case with intermediate value of $\nu^2$ in the same manner,
and we also find the minimum value of ${\mathcal{M}_{(Q)}^2 / H^2}$ is also affected by the values of $\Lambda$ and $\lambda$.
Namely, large $\Lambda>0$ and small $\lambda$ are preferred.

\begin{figure}[thbp]
\begin{center}
\includegraphics[width=80mm]{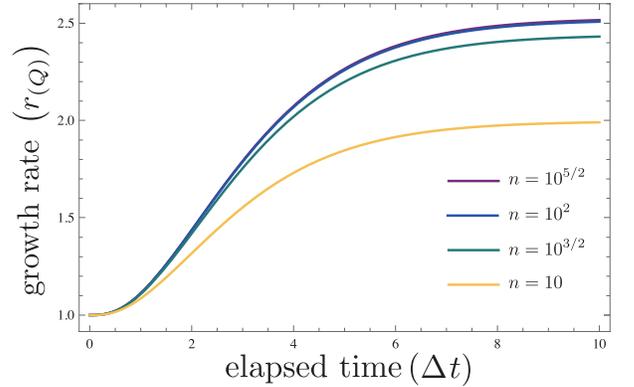}
\end{center}
\caption{
The evolutions of the scalar perturbation growth rates $r_{(Q)} := h_{(Q)}(t)/ h_{(Q)}(t_0)$ in terms of $n$ 
(solution (v) listed in TABLE \ref{table_stable_sol}).
In this numerical simulation, we set $\dot{h}_{(Q)}(t_0) =0$ and $\lambda =2.1$.
The yellow, green, blue and violet curves indicate the growth rate with $n=10, 10^{3/2}, 10^2$ and $10^{5/2}$, respectively. 
Note that the purple and blue curves are almost degenerated. 
The elapsed time is defined by $\Delta t := t- t_0$. 
In this solution, $\min_{a} {(\mathcal{M}_{(Q)}^2 / H^2)}$ approaches to $  {-3.621}$ for large perturbation mode.
}
\label{growth_scalar}
\end{figure}

Then, we demonstrate the growth of the scalar perturbation 
without condition (\ref{expanding_phase_stability}) by solving the equation of motion, numerically.
In FIG. \ref{growth_scalar}, we show the evolution of the scalar perturbation of solution (v) listed in TABLE \ref{table_stable_sol}.
In this case, $\min_{a} {(\mathcal{M}_{(Q)}^2 / H^2)}$ monotonically decreases as the perturbation mode becomes larger, 
and approaches to $  {-3.621}$ (see FIG. \ref{B-1_plot}).
Then, the growth rate of scalar perturbation is converged to 
$r_{(Q)}(\infty) := h_{(Q)}(\infty)/h_{(Q)}(t_0) \approx  {2.527}$ for large perturbation mode, 
where $t_0$ is a time at which $\mathcal{M}_{(Q)}^2=0$.
Since the growth rate can be suppressed as $r_{(Q)} \sim \mathcal{O}(1)$, 
the temporal tachyon instability may not provide serious effect to the background geometry.

\begin{figure}[thbp]
\begin{center}
\includegraphics[width=80mm]{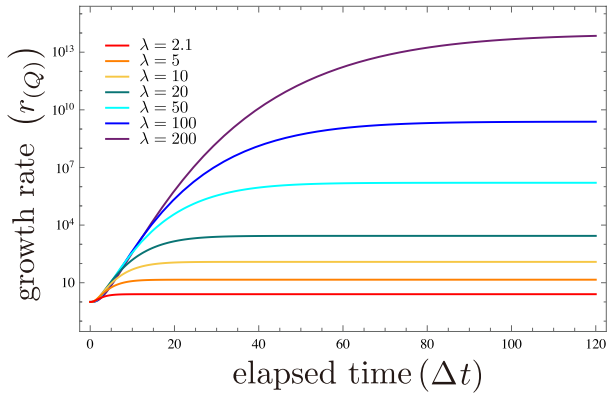}
\end{center}
\caption{
The evolutions of the scalar perturbation growth rates $r_{(Q)} := h_{(Q)}(t)/ h_{(Q)}(t_0)$ in terms of $\lambda$ 
(solution (v) listed in TABLE \ref{table_stable_sol}).
In this plot, we set $\dot{h}_{(Q)}(t_0) =0$ and $n=10^{5/2}$.
The red, orange, yellow, green, cyan, blue and violet curves indicate the growth rate with $\lambda=2.1, 5, 10, 20, 50, 100$ and $200$, respectively. 
The elapsed time is defined by $\Delta t := t- t_0$. 
}
\label{lambda_growth}
\end{figure}

Furthermore, we mention the relation between the scalar growth rate and the minimum value (\ref{MH_rate}).
To evaluate the relation,
we perform numerical calculation by setting various $\lambda$ with fixed perturbation mode.
Then, the evolutions of the scalar perturbation is shown in FIG. \ref{lambda_growth}.
Additionally, the detailed data of the asymptotic values of $r_{(Q)}$ in terms of 
$\lambda$ are shown in TABLE \ref{table_growth}.
From this result, we can conclude that the large value of $\left| \mathcal{M}_{(Q)}^2/H^2 \right|$ with $\mathcal{M}_{(Q)}^2 <0$ enhances the growth of scalar perturbation.
In other words, 
the scalar growth rate is amplified by choosing large value of $\lambda$ and small value of  $\Lambda>0$.
\begin{table}[tbhp]
\begin{center}
\begin{tabular}{cccccccccccccc}
\hline \hline
\rule[-1.5mm]{0mm}{5mm}&&~$\lambda$~~&~$\min \left( \mathcal{M}^2_{(Q)}/H^2\right)$~~&~$r_{(Q)}$~&\\
\hline
\rule[-1mm]{0mm}{4mm}&&~$2.1$~~&~$ {-3.621}$~~&~$ {2.527}$~&\\
\rule[-1mm]{0mm}{4mm}&&~$5$~~&~~$ {-1.317 \times 10^1}$~&~$ {1.437 \times 10^1}$~&\\
\rule[-1mm]{0mm}{4mm}&&~$10$~~&~~$ {-2.963 \times 10^1}$~&~$ {1.218 \times 10^2}$~&\\
\rule[-1mm]{0mm}{4mm}&&~$20$~~&~~$ {-6.255 \times 10^1}$~&~$ {2.749 \times 10^3}$~&\\
\rule[-1mm]{0mm}{4mm}&&~$50$~~&~~$ {-1.613 \times 10^{2}}$~&~$ {1.596 \times 10^6}$~&\\
\rule[-1mm]{0mm}{4mm}&&~$100$~~&~~$ {-3.259 \times 10^{2}}$~&~$ {2.418 \times 10^9}$~&\\
\rule[-1mm]{0mm}{4mm}&&~$200$~~&~~$ {-6.551 \times 10^{2}}$~&~$ {8.747 \times 10^{13}}$~&\\
\hline \hline 
\end{tabular}
\caption{
The detailed data of the relation between the value of $\lambda$ and asymptotic value the scalar growth rate.
In this table, we set $n=10^{5/2}$ (solution (v) listed in TABLE \ref{table_stable_sol}).}
\label{table_growth}
\end{center}
\end{table}

 {
When the accelerating expansion caused by a cosmological constant persists, 
the effect of the spacial curvature $K$ turns to be irrelevant.
Then, the analysis can be simplified into the case with $K=0$ and $\Lambda>0$.
In other  words, the spacetime can be approximated as the de Sitter solution 
at the late time of the evolution after bounce 
(c.f. cosmic no-hair theorem\cite{cosmic_no-hair}).
It should be noted that the detailed analysis of de Sitter spacetime stability
has been already performed in the papers\cite{dS_stability}.
In those paper, the authors indicated that the instability of scalar perturbation may be cured
if the background spacetime is the de Sitter solution.
It is worth mentioning that the quadratic actions in flat FLRW spacetime can be reproduced 
by simply taking a limit $K \to 0$ in (\ref{pert_action_tensor}) and (\ref{pert_action_scalar})\footnote{
Precisely, the function $X^{(n;l)}$ which is $\chi$ dependent part of the harmonics
is replaced by the spherical Bessel function when the spacial curvature $K$ is absent.
}.
}

\subsection{backreaction of the perturbation on background geometry}
\label{anisotropy_pert}
We further discuss the stability of bouncing solution by considering a backreaction of the perturbation. 
Especially, against anisotropic and homogeneous perturbation in closed FLRW spacetime.
Such perturbation modes can be derived by considering Bianchi type IX spacetime 
whose three-dimensional space is {homogeneous}, however, isotropy is not always hold.
The metric is given by
\begin{eqnarray}
ds^2 = -dt^2 + {a^2 \over 4} e^{2\beta_{ij}} \omega^i \omega^j \label{Bianchi_IX_metric}\,,
\end{eqnarray}
where, $a$ represents the scale factor, $\omega^i$ ($i=1,2,3$) is an invariant basis which is given by 
\begin{eqnarray}
\omega^1 &=& \sin x^3 \sin x^2\, dx^1 +\cos x^3 \,dx^2 \,, \notag \\
\omega^2 &=& -\cos x^3 \sin x^2\, dx^1 + \sin x^3\, dx^2 \,, \\
\omega^3 &=& \cos x^2 \,dx^1 + dx^3 \,. \notag
\end{eqnarray}
The traceless symmetric tensor $\beta_{ij}$ represents anisotropy.
Since Bianchi type IX space belongs to Bianchi class A, $\beta_{ij}$ can be diagonalized without loss of generality 
as follows\cite{Bianchi_class} : 
\begin{eqnarray}
\beta_{ij} = \mathrm{diag} (\beta_+ +\sqrt{3} \beta_-, \, \beta_+ -\sqrt{3} \beta_-,\,  -2\beta_+) \,.
\end{eqnarray} 
When $\beta_\pm =0$, the spacial isotopy is restored, namely, closed FLRW spacetime.
The basic equations are given by
\begin{eqnarray}
&&\dot{H} +3H^2
-{8 \over 3(3\lambda-1)} \left[ {8 \over a^5} {\partial V \over \partial a} +{3\mathcal{C}_{\mathrm{IX}} \over a^3}  \right] =0\,, \label{IX_eq1}
 \\
&&\ddot{\beta}_\pm +3H \dot{\beta}_\pm + {32 \over 3a^6} {\partial V \over \partial \beta_\pm} =0 \,, 
\label{eom_beta} \\
&& H^2 = {2 \over 3(3\lambda-1)} \left[ 3(\dot{\beta}_+^2 + \dot{\beta}_-^2) +{64 \over a^6} V(a,\beta_\pm) + {8 \mathcal{C}_{\mathrm{IX}} \over a^3} \right]\,, \notag \\ \label{IX_eq2}
\end{eqnarray}
where, $V(a, \beta_\pm):=-a^3 \mathcal{L}_P/128$ is a potential which is given by the spacial Ricci curvature terms
(in \cite{previous_2}, the explicit form is shown).
The equation (\ref{IX_eq1}) and (\ref{eom_beta}) correspond to 
the dynamical equations of the scale factor $a$ and the anisotropy $\beta_\pm$, respectively. 
As is the case with the case of FLRW spacetime, 
we obtain a Friedmann-like equation (\ref{IX_eq2}).
Due to integration with respect to time variable, a integration constant {$\mathcal{C}_{\mathrm{IX}}$} appears.
For simplicity, we consider the case with $\mathcal{C}_{\mathrm{IX}} =0$.

Assuming $|\beta_\pm| \ll 1$, the potential $V$ is reduced into 
\begin{eqnarray}
V(a, \beta_\pm) \approx  U_0(a) + U_{2}(a) (\beta_+^2 + \beta_-^2)\,,
\end{eqnarray}
where, $U_{0}(a)$ and $U_{2}(a)$ are defined by
\begin{eqnarray}
U_{0}(a) &:=& -{ 3a^4 \over 64} \left[ 1 - {\Lambda \over 3}a^2 -{g_r \over 3a^2} -{g_s \over 3a^4}  \right] \,, \\ \notag \\
U_{2}(a) &:=& {3a^6\over 64} \bigg[ {8 \over a^2} +{16 \over 3 a^4} \left(12g_3 -g_r\right)  \notag \\
&&~~~~~~~
 +{8 \over a^6} \Big\{ 48(g_{56} +g_8) -g_s \Big\} \bigg] \,.
\end{eqnarray}
It is worth mentioning that $U_0$ and $U_2$ are related to the potential in FLRW spacetime (\ref{FLRW_potential_form}) and 
the squared effective mass $\mathcal{M}^2 := \mathcal{G}/\mathcal{F}$ of $n=3$ tensor mode as follows : 
\begin{eqnarray}
&&\mathcal{U}(a) = -{64 \over 3(3\lambda-1)a^4}U_0(a)\,, \\
&&\mathcal{M}_{(G)}^2\Big|_{n=3} = {64 \over 3a^6} U_2(a) \,.
\end{eqnarray}
One may notice that the equation of motion for the tensor perturbation with $n=3$ is reproduced 
if $\beta_\pm$ is replaced into $h_{(G)}$ in (\ref{eom_beta}) (see (\ref{tensor_n3}) and (\ref{G_eom})).
In other words, this perturbation modes include the homogeneous and anisotropic perturbation in closed FLRW spacetime.
Then, the equation (\ref{IX_eq2}) can be rewritten as follows :
\begin{eqnarray}
{1 \over 2}\dot{a}^2 +\mathcal{U}(a) \approx {2 a^2 \over 3\lambda -1} \left[ E_{\beta_+}(a,\beta_+) + E_{\beta_-}(a,\beta_-) \right]\,,
\notag \\
\label{H_IX}
\end{eqnarray}
where, 
\begin{eqnarray}
E_{\beta_\pm}(a, \beta_\pm) &:=& {1 \over 2} \left[ \dot{\beta}_\pm^2 + \mathcal{M}_{(G)}^2\Big|_{n=3} \beta_\pm^2 \right]\,.
\label{E_beta}
\end{eqnarray}
Since we impose the positivity of the tensor squared mass $\mathcal{M}^2_{(G)}$
to stabilize the perturbation, $E_{\beta_\pm}$ always takes positive value.
From (\ref{H_IX}), one can see that the scale factor is regarded as a particle with
energy $2a^2(E_{\beta_+} + E_{\beta_-})/(3\lambda-1)$ in potential $\mathcal{U}(a)$.
Namely, the possible range for the scale factor is broaden due to an {\it anisotropic energy} $E_{\beta_\pm}$.

\begin{figure}[htbp]
\begin{center}
\includegraphics[width=80mm]{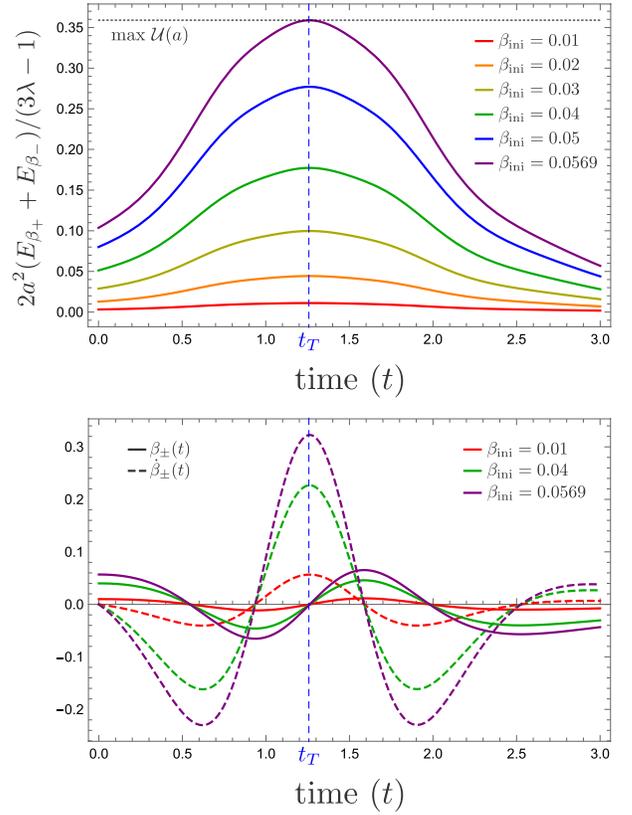}
\end{center}
\caption{
The evolutions of anisotropy of $\mathcal{B}^{[1]}_{BC}$ (solution (i) listed in TABLE \ref{table_stable_sol}).
In the top figure, the evolution of anisotropic energies are shown.
The initial conditions are given by $a_{\mathrm{ini}}=1.5a_T \approx 2.784$, $\beta_\pm=\beta_{\mathrm{ini}}$ and $\dot{\beta}_\pm=0$.
The coupling constant $\lambda$ is set to be unity.
The red, orange, yellow, green, blue, purple curves correspond to the anisotropic energy
with initial anisotropy $\beta_{\mathrm{ini}} = 0.01, 0.02, 0.03, 0.04, 0.05$ and $0.0569$, respectively.
The maximum value of the potential in FLRW spacetime $\mathcal{U} \approx 0.359$
is denoted by the gray dotted line.
In the bottom figure, 
the solid and dashed curves indicate the evolutions of $\beta_\pm$ and $\dot{\beta}_\pm$.
The red, green and purple curves correspond to the evolutions with initial condition 
$\beta_{\mathrm{ini}} = 0.01, 0.04$ and $0.0569$, respectively.
}
\label{backreaction_BC}
\end{figure}

Then, we examine the backreaction on the singularity-free solutions, especially $\mathcal{B}^{[1]}_{BC}$.
This type of bouncing solutions realizes singularity avoidance due to the potential barrier $\mathcal{U} \geq 0$
between $a_{BC}$ and $a_T$ (see FIG.\ref{pot_FLRW_K+}).
However, the cosmological bounce at $a_T$ may be spoiled 
if the backreaction from anisotropic perturbation is considered.
Namely, the energy for the scale factor is lifted up to $2a^2(E_{\beta_+} + E_{\beta_-})/(3\lambda-1)$ 
due to anisotropic effect, 
and then, the potential barrier can be overleaped 
if the anisotropic energy exceeds the local maximum value of the potential $\mathcal{U}$.
In FIG. \ref{backreaction_BC}, we show the typical example based on the solution (i) listed in TABLE \ref{table_stable_sol}.
In this analysis, we set $\lambda=1$ and the initial conditions are given by 
$a_{\mathrm{ini}}=1.5a_T \approx 2.784$, $\beta_\pm=\beta_{\mathrm{ini}}$ and $\dot{\beta}_\pm=0$.
In the top figure, the evolutions of the anisotropic energies are shown.
The anisotropic energy $2a^2(E_{\beta_+} + E_{\beta_-})/(3\lambda-1)$ takes maximum value
at the bouncing time $t=t_{T} \approx 1.261$.
If the initial anisotropy exceeds a critical value $\beta_{\mathrm{crit}} \approx 0.0569$, 
the universe results in big crunch by overleaping the potential barrier 
(the purple curves in the top figure of FIG. \ref{backreaction_BC}). 

We further mention the dynamics of the anisotropy $\beta_\pm$ and $\dot{\beta}_\pm$ 
(the bottom figure of FIG. \ref{backreaction_BC}).
One may notice that the oscillating amplitudes of $\beta_\pm$ are almost invariant throughout bounce,
while those of $\dot{\beta}_\pm$ are enhanced whose maximum amplitudes reach up to $10^{-1}$ order.
It is not quite unnatural because the dynamics of anisotropic perturbation 
is approximately governed by the following equation : 
\begin{eqnarray}
\ddot{\beta}_\pm \approx -\mathcal{M}^2_{(G)} \beta_\pm\,.
\end{eqnarray}
Around the bouncing point, the term including Hubble parameter can be ignored. 
Then, the oscillating frequency is naively given by 
$\omega \approx \mathcal{M}_{(G)} = \sqrt{\mathcal{G}_{(G)}/\mathcal{F}_{(G)}}$.
As a result, the amplitudes of $\dot{\beta}_\pm$ is approximately estimated by
$|\dot{\beta}_\pm| \approx \mathcal{M}_{(G)}  |{\beta}_\pm|$.
Since $\mathcal{M}^2_{(G)} \propto a^{-6}$ for small scale factor, 
the large anisotropic energy at bouncing point is induced if we consider small bouncing radii.  
Thus, we conclude that the backreaction on the bouncing universe with large bouncing radii tend to be small.

The backreactions from the other perturbation mode are unclear.
However, it is natural to consider that the most symmetric spacetime corresponds to
the lowest energy state.
Then, one may speculate that the other perturbation modes also lift up the energy for the scale factor  
like the case of the anisotropic perturbation as (\ref{H_IX}) and (\ref{E_beta}). 

We also would like to point out that the oscillating universe obtained by $\mathcal{B}^{[1]}_{\mathcal{O}}$ with
$a_{\mathrm{min}} \leq a_{\mathrm{ini}} \leq a_{\mathrm{min}}$ can possibly evolve into macroscopic universe.
In this type of potential {$\mathcal{U}(a)$}, the oscillating and bouncing solutions are separated by 
a potential barrier between $a_{\mathrm{max}}$ and $a_T$ (see FIG.\ref{pot_FLRW_K+}).
However, if the potential barrier is sufficiently small, 
it is possible that the initial oscillating era shifts into  {accelerating} expanding phase via energy induced by perturbations.
In fact, the similar evolution of the universe is found when the spacetime anisotropy is large\cite{previous_2}.

%======================================%
%<<<<<<<<<<<< SECTION V  >>>>>>>>>>>>>>%
%======================================%
%%%%%%%%%%%%%%%%%%%%%%%%%%%%%%%%%%%%%%%%%%%
%%%%%%%%%%%%%%%%%%%%%%%%%%%%%%%%%%%%%%%%%%%
%%%%%%%%%%%%%%%%%%%%%%%%%%%%%%%%%%%%%%%%%%%
%%%%%%%%%%%%%%%%%%%%%%%%%%%%%%%%%%%%%%%%%%%
\section{conclusion}
\label{conclusion}
To avoid big-bang singularity at the beginning of universe, 
it is essential to consider the situation 
which null energy condition is violated, i.e., $p+\rho = (1+w)\rho <0$.
Based on HL theory, the higher spacial curvatures in action possibly 
behave as such exotic matters.
Thus, one find singularity-free cosmological solutions, such as bouncing universe 
in non-flat FLRW spacetime.
However, it is natural to consider the effective exotic matters which violate null energy condition
destabilize spacetime.

In this paper, we investigate the stabilities of bouncing solutions 
via perturbation analysis around non-flat FLRW spacetime.  
Employing { (pseudo-)spherical} harmonic functions, both tensor and scalar perturbations can be decomposed 
into each $(n;l,m)$ modes.
Then, perturbed actions for tensor and scalar modes at quadratic order are reduced into (\ref{pert_action_tensor}) and (\ref{pert_action_scalar}), respectively. 
Note that the integration constant $\mathcal{C}$ induced by the lack of local Hamiltonian constraint 
does not affect to the quadratic action, 
however, background dynamics is influenced, 
i.e., dust-like additional term is joined in the Friedmann-like equations.
In our analysis, the integration constant is set to zero, for simplicity.
Thus, the result may be slightly changed if we consider non-zero $\mathcal{C}$, 
i.e., the spacetime stability around bouncing point.

In order to avoid ghost instabilities, we must require the coefficients 
of kinetic terms in quadratic action to be positive, namely, 
$\mathcal{F}_{(G)} \geq 0$
and $\mathcal{F}_{(Q)} \geq 0$ for any $a>0$ and perturbation mode $\nu^2$.
Since $\mathcal{F}_{(G)} =1$, tensor perturbations do not show ghost instability.
On the other hand, the condition for ghost avoidance in scalar perturbation
is expressed in terms of $\lambda$ :
 \begin{eqnarray}
\begin{cases}
\lambda \geq 1\,, & \text{for}~K=+1 \,, \\[3mm]
\lambda > 2\,, & \text{for}~K=-1 \,. 
\end{cases}
\end{eqnarray}
Note that in flat FLRW spacetime or Minkowski spacetime, the stability condition for scalar perturbation 
is given by $\lambda >1$ and scalar degree of freedom vanishes when $\lambda =1$.
It is known that there is no smooth connection between $\lambda>1$ and $\lambda=1$ because of 
strong coupling problem.
However, in closed FLRW case, we can take smooth limit $\lambda \to +1$ without any singular 
behavior, 
and then, the scalar perturbation can propagate even if the case with $\lambda =1$ is considered. 
Thus, the limit $\lambda \to 1$ does not mean GR is restored.
The dissimilarity is due to the gauge structure mentioned in Appendix \ref{sec_gauge_fixing}.
 
We further consider the positivities of $\mathcal{G}_{(G)}$ and $\mathcal{G}_{(Q)}$
for any perturbation mode $(n; l, m)$.
In order to stabilize the perturbation modes with large $n$, 
the following conditions must be satisfied : 
\begin{eqnarray}
g_8 \geq 0 \,,~g_8 \geq 8g_7/3\,.
\end{eqnarray}
Although it is possible that $\mathcal{G}_{(G)} \geq 0$ for any $a>0$ and viable $\nu^2$,  
the negativity of $\mathcal{G}_{(Q)}$ cannot be avoided in infrared regime, i.e., 
there must be exist $a_{\mathrm{crit}}$ 
at which any one of scalar perturbation mode turns from $\mathcal{G}_{(Q)}>0$ to $\mathcal{G}_{(Q)}=0$.
This result is consistent with infrared instability of scalar graviton in flat background.
Note that the negative value of $\mathcal{G}_{(Q)}$ does not always mean the instability of the scalar perturbation.
Then we have investigated the dynamics of perturbations in bouncing universe via equations of motion. 

In contracting phase, perturbations are possibly amplified by the negative sign of Hubble term.
To suppress the instabilities, the following condition must be satisfied :
\begin{eqnarray}
\mathcal{M}^2_{(G)} \gtrsim H^2 \,,~ \mathcal{M}^2_{(Q)} \gtrsim H^2  &\text{for} ~~a_T < a \leq a_{\mathrm{ini}}\,.
\end{eqnarray}
In infrared regime, the squared effective masses of scalar perturbations must be negative
for any choice of the coupling constants $g_i$.
Thus, we impose the following condition in order to overcome the effect of $\mathcal{M}_{(Q)}^2<0$ :
\begin{eqnarray}
\left| \mathcal{M}^2_{(Q)} \right| \lesssim H^2 &\text{for} ~~ a \gtrsim a_{\mathrm{crit}}\,,
\label{IR_stability}
\end{eqnarray}
Thus, the stable bouncing solutions are limited to the case with $\Lambda >0$.
If the condition (\ref{IR_stability}) is violated, scalar perturbation is amplified,
which means the tachyon instability is occurred.
Since $ \left |\mathcal{M}_{(Q)}^2 \right|$ decreases as $a^{-2}$ at infrared regime, 
the period for the scalar instability is temporal.
The growth rates of the scalar perturbations are related with the minimum value of $\mathcal{M}_{(Q)}^2 /H^2$.
Then, small $\lambda -1 >0$ and/or large $\Lambda>0$ are preferred
to suppress the growth of scalar perturbations.
It may be interesting to estimate permissible growth rate of scalar perturbation by
referring observational cosmological data. 
Then, one can derive further constraints to the values of coupling constants.

Additionally, we have investigated backreaction from perturbation on background geometry,
especially, against anisotropic perturbation in $\mathcal{B}_{BC}^{[1]}$ type solution.
Considering Bianchi type IX spacetime with small anisotropy, 
the modified Friedmann-like equation including backreaction from the anisotropic perturbation can be derived as (\ref{H_IX}).
It is found that the energy for the scale factor is lifted up by anisotropic perturbation.
Thus, the universe can evolve into the singularity if the potential barrier between $a_{BC}$ and $a_{T}$
is sufficiently small.
We also pointed out that the anisotropic energy tend to be large if the bouncing radius is small,
because the oscillating amplitudes of $\dot{\beta}_\pm$ are enhanced.

We would like to stress that the bouncing solutions in open FLRW spacetime 
tend to be unstable for the following reasons : 
(i) the stability conditions for tensor perturbation basically contradicts to bouncing conditions.
Intuitively, $g_r>0$ and $g_s>0$ are preferred to satisfy $\mathcal{G}_{(G)} \geq 0$ 
for any $a>0$ and viable $\nu^2$.
However, we cannot find any types of bouncing solution in open FLRW spacetime
under such a condition.
Thus, a certain level of tuning of the other coupling constants is required.
(ii) in infrared regime,  the Hubble friction which is significant to suppress tachyon 
instabilities in scalar perturbation tend to be weak.
In closed FLRW spacetime, the minimum value of $\mathcal{M}_{(Q)}^2 /H^2$
can be converged to zero if $\lambda$ approaches to unity via renormalization group flow. 
However, in open FLRW spacetime, 
we have to constrain $\lambda>2$ to avoid ghost instabilities.

Our conclusion is that we have shown that 
non-singular cosmological solutions in non-flat FLRW spacetime can be stable against tensor and scalar perturbation,
at least at the linear level.
Since projectable HL theory is proved to be truly renormalizable in perturbation approach,
we can calculate the values of the coupling constants via beta functions from renormalization group, in principle. 
Thus, it may be possible that the beginning of our Universe can be predicted 
based on well-known perturbative quantization approach.

In our case, i.e., HL theory under projectability condition, 
it is indispensable to consider  {accelerating} expanding phase after bounce
in order to suppress the effect of negative squared mass of scalar perturbation in infrared regime.
We speculate that the scalar instabilities in infrared regime is
the nature of projectable HL gravity theory.
This infrared pathological behavior is conceivably resolved by considering 
extended theory, i.e., non-projectable HL gravity whose scalar graviton
can be stable at least in Minkowski spacetime.

We additionally mention that infrared limit of non-projectable HL theory 
(i.e., without higher spacial curvatures) is included within the framework of Horndeski theory
which is the general theory of ghost-free scalar-tensor gravity\cite{Horndeski_theory}.
Based on Horndeski theory, it turns out that any non-singular cosmological solution is unstable\cite{no-go_Horndeski}.
On the other hand, it is worth mentioning that the no-go theorem for the stable non-singular solution can be violated 
if the extended theory including higher spacial curvatures is considered\cite{bounce_EFT}.
Thus, in view of the situation, it should be interesting to investigate the stability of bouncing solutions 
with higher spacial curvatures based on non-projectable HL theory 
as a special case of extended Horndeski scalar-tensor gravity\cite{forthcoming_paper}.

%======================================%
%<<<<<<<<<<<< Acknowledgments >>>>>>>>>>>>%
%======================================%
\section*{Acknowledgments}
The authors would like to thank Kei-ichi Maeda, Shuntaro Mizuno and Katsuki Aoki for valuable comment and discussions.

%%%%%%%%%%%%%%%%%%%%%%%%%%%%%%%%%%%%%%%%%%%%%%%%%%

%======================================%
%<<<<<<<<<<<< APPENDIX >>>>>>>>>>>>%
%======================================%
%%%%%%%%%%%%%%%%%%%%%%%%%%%%%%%%%%%%%%%%%%%%%%%%%%
%%%%%%%%%%%%%%%%%%%%%%%%%%%%%%%%%%%%%%%%%%%%%%%%%%

 \appendix

 \section{Spherical and Pseudo-Spherical Harmonics}
 \label{app_3-harmonics}
 The detailed discussion about the tensor (pseudo-) spherical harmonics has 
 already performed in the papers \cite{Ynlm_ref} and \cite{SandPS}.
 In this section, we give the correspondence between 
 our definition and those of above references.
 
 \subsection{Tensor spherical harmonics on two-sphere}
 Before considering the three-dimensional case, 
 we introduce spherical harmonics on unit two sphere whose metric $s_{AB}$ is given by
 \begin{eqnarray}
ds_{(2)}^2 = d\theta^2 + \sin^2 \theta d\phi^2 \,.
 \end{eqnarray}
 Then, the scalar spherical harmonics $Y^{(lm)}(\theta,\phi)$ is given by
 \begin{eqnarray}
 Y^{(lm)} &=& (-1)^{(m+|m|)/2}  \sqrt{ {(2l+1) (l-|m|)! \over 4\pi (l+|m|)!} } \notag \\
 &&\times P_l^{|m|} (\cos \theta) e^{im \phi} \,, 
 \end{eqnarray}
 where, $P_l^{|m|} (\cos \theta) $ is a Legendre polynomial 
 whose Rodrigues's formula is given by
 \begin{eqnarray}
 P^m_n (x) ={(-1)^n \over 2^n n!} (1-x^2)^{m/2} {d^{n+m} \over dx^{n+m}}(1-x^2)^n \,.
 \end{eqnarray} 
 where, degrees $l, m \in \Z$ are constrained by $0 \leq |m| \leq l$.
 
 The vector spherical harmonic functions are classified into two classes. 
 Since the vector quantity can be decomposed into gradient part and rotational part,
 we define the gradient of the scalar harmonics $\psi_A^{(lm)}$ and 
 the dual of the gradient  $\phi_A^{(lm)}$ :
 \begin{eqnarray}
 \psi_A^{(lm)} &:=& \mathcal{D}_A Y^{(lm)} 
 \,, \\
 \phi_A^{(lm)} &:=& \epsilon_A^{~B} \mathcal{D}_B Y^{(lm)} 
 \,, 
 \end{eqnarray}
where, 
$\mathcal{D}_A$ denotes a covariant derivative on two sphere, $\epsilon_{AB}$ is Levi-Civita tensor on two-sphere.
 Note that $\psi_A$ and $\phi_A$ posses even and odd parity, respectively.
 
 The tensoral ones are classified into three-types :
 $\eta_{AB}$ is a trace part, which is proportional to two-metric. 
 $\psi_{AB}$ and $\phi_{AB}$ are traceless with even and odd parity,
 respectively.
 The explicit forms are given by
 \begin{eqnarray}
 \eta^{(lm)}_{AB} &:=& Y^{(lm)} s_{AB}
  \,, \\
 \psi^{(lm)}_{AB} &:=& \mathcal{D}_A \mathcal{D}_B Y^{(lm)} +{l(l+1) \over 2} Y^{(lm)} s_{AB}\,, \\
 \phi^{(lm)}_{AB} &=& {1 \over 2} \left[ \mathcal{D}_{A} \phi_B^{(lm)} + \mathcal{D}_{B} \phi_A^{(lm)} \right] \,.
 \end{eqnarray}
 
 \subsection{Tensor harmonics on three (pseudo-)sphere}
 We consider the harmonics on unit three sphere and unit three pseudo-sphere whose metric
 $\ha{\gamma}_{ij}$ and $\ch{\gamma}_{ij}$ are given by 
\begin{eqnarray}
d \ha{\ell}^2 &=& d \chi ^2 + \sin^2 \chi ( d \theta^2 + \sin^2 \theta \, d \phi^2 )\,, \\
d \ch{\ell}^2 &=& d \chi ^2 + \sinh^2 \chi ( d \theta^2 + \sin^2 \theta \, d \phi^2 )\,.
\end{eqnarray}
 To construct the harmonics for each cases, we require that
 (i) the harmonic function ${\bf Y}$ has the eigenvalues of Laplace-Beltrami operator,  
 (ii) the orthonormality is satisfied. 
 Then, the functions defined on three sphere and three pseudo-sphere can be expanded by each ($n,l,m$) modes of harmonics.
For spherical case, $n \geq 1$ is discrete natural number which constrain $ 0 \leq l \leq n-1$.
On the other hand, for pseudo-spherical case, $n \geq 1$ is defined as a continuous number.
 
We give the explicit form of the scalar spherical harmonics $\ha{Y}^{(n;lm)}$ :
 \begin{eqnarray}
\ha{Y}^{(n;lm)} (\chi,\theta, \phi) &:=& 
\ha{X}^{(n;l)}(\chi) Y^{(lm)}(\theta,\phi) \,,  
\\ \notag \\
\ha{X}^{(n;l)}(\chi) &:=&  \sqrt{ {2 \over \pi}  } \left[ \prod_{0 \leq k \leq l} {1 \over n^2 -k^2} \right]^{1/2} \notag \\
&& \times \sin^l \chi  \, { d^{(l+1)} \over d(\cos \chi)^{l+1}} \cos \left( n\chi \right)\,. \notag \\
\label{eq_def_Xnl}
\end{eqnarray}
Note that $\ha{X}^{(n;l)}$ is expressed in terms of Gegenbauer (ultraspherical) polynomials
which is a generalization of Legendre polynomials\cite{Ynlm_ref,AandS}.
The eigenvalues of Laplace-Beltrami operator are given by
\begin{eqnarray}
\ha{\mathpzc{D}}^2 \, \ha{Y}^{(n;lm)} = -(n^2-1) \ha{Y}^{(n;lm)}\,,
\end{eqnarray}
where, $\ha{\mathpzc{D}_i}$ denotes a covariant derivative in terms of $\ha{\gamma}_{ij}$.
One can confirm that the above harmonics satisfy orthonormality : 
\begin{eqnarray}
\left\langle \ha{Y}^{(n;lm)}, \ha{Y}^{(n';l'm')} \right\rangle
&=& \delta_{(n,n')} \delta_{(l,l')} \delta_{(m,m')} \,,
\end{eqnarray} 
where, we define the internal product on three-sphere as 
\begin{eqnarray}
\left\langle \ha{{\bf Y}}_1, \ha{{\bf Y}}_2  \right\rangle := \int_{0}^{\pi} \!\!\!\! d\chi 
\int_{0}^{\pi} \!\!\!\! d\theta \int_{0}^{2\pi} \!\!\!\!\!\! d\phi ~\sqrt{\ha{\gamma}}\, \ha{\bf Y}_1\cdot \ha{\bf Y}_2 \,,
\label{IP_spherical}
\end{eqnarray}
a symbol $\cdot$ denotes a contraction of tensor indices.
  
The pseudo-spherical harmonics $\ch{Y}^{(n;lm)}$ can be 
derived from the three-spherical ones by considering analytic continuation, 
namely $\chi \to i\chi$ and $n \to in$\cite{Lifshitz_Khalatnikov,SandPS} :
\begin{eqnarray}
\ch{Y}^{(n;lm)} &=& \ch{X}^{(n;l)}(\chi)\, Y^{(lm)}(\theta,\phi) \,, \\
\ch{X}^{(n;l)} &=&  \sqrt{ {2 \over \pi} } \left[\prod_{0 \leq k \leq l} {1 \over n^2 +k^2}\right]^{1/2} \notag \\
&& \times \sinh^l \chi\, {d^{(l+1)} \over d(\cosh \chi)^{l+1}} \cos \left( n\,\chi \right) \,. \notag \\
\end{eqnarray}
The eigenvalues of Laplace-Beltrami operator on unit three pseudo-sphere are given by
\begin{eqnarray}
\ch{\mathpzc{D}}^2 \ch{Y}^{(n;lm)} = -(n^2 +1) \ch{Y}^{(n;lm)} \,,
\end{eqnarray} 
where, $\ch{\mathpzc{D}_i}$ denotes a covariant derivative in terms of $\ch{\gamma}_{ij}$.
The orthonormality is also satisfied if we define the internal product on three unit pseudo-sphere
as follows :
\begin{eqnarray}
\left\langle \ch{ {\bf Y} }_1, \ch{{\bf Y}}_2  \right\rangle :=\lim_{L \to \infty} {\pi \over L}
 \int_{0}^{L} \!\!\!\! d\chi 
\int_{0}^{\pi} \!\!\!\! d\theta \int_{0}^{2\pi} \!\!\!\!\!\! d\phi ~\sqrt{\ch{\gamma}}\, \ch{\bf{Y}}_1 \cdot \ch{ {\bf Y}}_2 \,, \notag \\
\label{IP_pseudo-spherical}
\end{eqnarray}
Then, the orthonormality of pseudo-spherical harmonics is given by
\begin{eqnarray}
\left\langle \ch{Y}^{(n;lm)}, \ch{Y}^{(n';l'm')} \right\rangle
&=& \delta(n-n') \delta_{(l,l')} \delta_{(m,m')} \,, \notag \\
\end{eqnarray} 
In order to unify the discussion of both cases,  
we define the eigenvalues of spherical and pseudo-spherical harmonics as follows 
\begin{eqnarray}
&&\nu^2 := 
\begin{cases}
n^2 -1\,,~ n\in \N &\text{for}~ K=1 \\
n^2 +1\,,~ n \in \R &\text{for}~ K=-1 \\
\end{cases}\,. \label{eigan_nu}
\end{eqnarray}
In what follows, 
we abbreviate the superscripts $\, \ha{} \,$ and $\,\ch{} \,$ if not otherwise specified.

 \subsubsection{scalar type}
 We introduce the scalar type harmonics which contribute to the scalar perturbation. 
 Since the scalar quantities have already been introduced in previous part, 
 we focus only on the vector and tensor quantities.
 
The vector quantities $C_i$ are defined by
\begin{eqnarray}
C^{(n;lm)}_i = {\mathpzc{D}}_i \, {Y}^{(n;lm)} \,.
\end{eqnarray}
The tensor quantities are classified into two kinds : 
\begin{eqnarray}
{C}^{(n;lm)}_{ij} &=& {\mathpzc{D}}_i {\mathpzc{D}}_j {Y}^{(n;lm)} +{\nu^2 \over 3} {Y}^{(n;lm)} {\gamma}_{ij} \,, \\
{D}^{(n;lm)}_{ij} &=& {Y}^{(n;lm)} {\gamma}_{ij} \,.
\end{eqnarray}
Namely, $C_{ij}$ and $D_{ij}$ assume the traceless and trace parts, respectively.
Considering the internal products, the normalized scalar type harmonics are defined by
\begin{eqnarray}
{Q}^{(n;lm)} &:=& {Y}^{(n;lm)} \,, \\
{Q}^{(n;lm)}_i &:=& \nu^{-1} {C}_i^{(n;lm)} \,, \\
{Q}^{(n;lm)}_{ij} &:=& { 1 \over \sqrt{3}  } {D}_{ij}^{(n;lm)} \,, \\
{P}^{(n;lm)}_{ij} &:=& \left[{2 \over 3}\nu^2 \left(\nu^2-3 K \right) \right]^{-1/2} {C}_{ij}^{(n;lm)} \,. 
\end{eqnarray}
When we consider spherical (pseudo-spherical) case, the spacial curvature takes $K=1$ $(K=-1)$. 

\subsubsection{vector type}
The vector type harmonics contribute to the transverse modes of the metric perturbation.
Namely, the divergences of these harmonics are vanished.

The vector quantities includes two types of harmonics.
One is odd parity mode ${A}_i$ whose explicit form is given by
\begin{eqnarray}
{A}^{(n;lm)}_i &=& \left( 0,~ f(\chi) \, {X}^{(n;l)} \phi^{(lm)}_A \right)\,.
\end{eqnarray}
where, the function $f(\chi)$ is defined in (\ref{def_f}).
Then, the eigenvalues are given by
\begin{eqnarray}
\mathpzc{D}^2 A^{(n;lm)}_i = -\left( \nu^2 -K \right) A^{(n;lm)}_i \,.
\end{eqnarray}
The other is even parity mode ${B}_i$ whose explicit form is given by
\begin{eqnarray}
{B}^{(n;lm)}_i &=& -{\epsilon}_{i}^{~jk} {\mathpzc{D}}_j {A}^{(n;lm)}_k\,,
\end{eqnarray}
where, ${\epsilon}_{ijk}$ is Levi-Civita symbol associated with ${\gamma}_{ij}$.
The eigenvalues are as same as odd ones.

The tensor quantities can be constructed by taking symmetrized gradient of each vector quantities :
\begin{eqnarray}
{A}^{(n;lm)}_{ij} &=& {1 \over 2} \left[ {\mathpzc{D}}_i {A}^{(n;lm)}_{j} + {\mathpzc{D}}_j {A}^{(n;lm)}_{i} \right]\,, \\
{B}^{(n;lm)}_{ij} &=& {1 \over 2} \left[ {\mathpzc{D}}_i {B}^{(n;lm)}_{j} + {\mathpzc{D}}_j {B}^{(n;lm)}_{i} \right]\,.
\end{eqnarray}
The eigenvalues are given by
\begin{eqnarray}
\mathpzc{D}^2 A^{(n;lm)}_{ij} = -\left( \nu^2 -5K \right) A^{(n;lm)}_{ij} \,.
\end{eqnarray}
The even parity modes have the identical eigenvalue as odd ones. 
Then, the normalized vector type harmonics are defined by
\begin{eqnarray}
{S}^{(n;lm)}_{(o)i} &:=& \left[ l(l+1)\right]^{-1/2} {A}_{i}^{(n;lm)} \,, \\
{S}^{(n;lm)}_{(e)i} &:=& \left[ l(l+1) \left(\nu^2+K\right) \right]^{-1/2} {B}_{i}^{(n;lm)} \,, \\
{S}^{(n;lm)}_{(o)ij} &:=& \left[ {l(l+1) \over 2} \left(\nu^2-3K \right)  \right]^{-1/2} {A}_{ij}^{(n;lm)} \,, \\
{S}^{(n;lm)}_{(e)ij} &:=& \left[ {l(l+1) \over 2} \left(\nu^2-3K \right)\left(\nu^2+K \right)  \right]^{-1/2} {B}_{ij}^{(n;lm)} \,. \notag \\
\end{eqnarray}

\subsubsection{tensor type}
The tensor type harmonics contribute to the transverse-traceless mode of perturbation.
Thus, we find only tensor quantities in this type.
As is the case with the vector type harmonics, there are odd and even parity mode.
The odd parity modes ${E}^{(n;lm)}_{ij}$ are given by 
\begin{eqnarray}
{E}^{(n;lm)}_{\chi \chi} &=& 0 \,,~ \\
{E}^{(n;lm)}_{\chi A} &=& {X}^{(n;l)} \phi^{(lm)}_A  \,,~ \\
{E}^{(n;lm)}_{\chi A} &=& \left[{2 \over (l + 2) (l - 1)}\right] {d \over d \chi} \left[ f(\chi)^{2} \, {X}^{(n;l)} \right] \phi^{(lm)}_{AB}  \,, \notag \\
\end{eqnarray}
The eigenvalue equation is given by 
\begin{eqnarray}
\mathpzc{D}^2 {E}^{(n;lm)}_{ij}  =-(\nu^2 -2K) {E}^{(n;lm)}_{ij} \,.
\end{eqnarray}
Those of even parity ${F}^{(n;lm)}_{ij}$ are given by
\begin{eqnarray}
{F}^{(n;lm)}_{ij} = {1 \over 2} \left[ {\epsilon}_{i}^{~pq} {\mathpzc{D}}_{p} {E}^{(n;lm)}_{qj} + {\epsilon}_{j}^{~pq} {\mathpzc{D}}_{p} {E}^{(n;lm)}_{qi} \right]\,.
\end{eqnarray}
The eigenvalues are as same as odd ones.
Then, the normalized tensor type harmonics are defined by
\begin{eqnarray}
{G}^{(n;lm)}_{(o)ij} &:=& \left[ {2l(l+1)  \over (l+2)(l-1)} \nu^2 \right]^{-1/2} {E}_{ij}^{(n;lm)} \,, \\
{G}^{(n;lm)}_{(e)ij} &:=& \left[ {2l(l+1) \over (l+2)(l-1)} \nu^2 \left( \nu^2+K\right) \right]^{-1/2} {F}_{ij}^{(n;lm)} \,. \notag \\
\end{eqnarray}

We shall summarize the properties of the spherical and pseudo-spherical harmonics 
in TABLE \ref{SandPS_table}.
\begin{table*}[tbhp]
\begin{center}
\begin{tabular}{cc|c|c|c|c|c|cc}
\hline \hline
\rule[-1.7mm]{0mm}{5mm}&~type~~&~$\mathbf{Y}$~~&~parity~&~trace~&~divergence~~&~eigenvalues~~&~viable degrees~~&~\\
\hline \hline 
\rule[-3mm]{0mm}{8mm}&
\multirow{10}{*}{scalar} &${Q}^{(n;lm)}$ & even & N/A & N/A & $-\nu^2$ & $n \geq 1$ \\ \cline{3-9}
\rule[-5mm]{0mm}{12mm}&&${Q}^{(n;lm)}_i$ & even & N/A & $\bigcirc$ & $-\nu^2+2K$ & 
$
\begin{cases}
n \geq 2  &\mathrm{for}~K=1 \\
n \geq 1  &\mathrm{for}~K=-1 
\end{cases}
$ \\ \cline{3-9}
\rule[-3mm]{0mm}{8mm}&&${Q}^{(n;lm)}_{ij}$ & even & $\bigcirc$ & $\bigcirc$ & $-\nu^2$ & $n \geq 1$ \\ \cline{3-9}
\rule[-5mm]{0mm}{12mm}&&${P}^{(n;lm)}_{ij}$ & even & $\times$ & $\bigcirc$ & $-\nu^2+6K$ & 
$
\begin{cases}
n \geq 3  &\mathrm{for}~K=1 \\
n \geq 1  &\mathrm{for}~K=-1 
\end{cases}
$ \\ \hline
\rule[-3mm]{0mm}{8mm}&
\multirow{3}{*}{vector}&${S}^{(n;lm)}_i$ & odd and even & N/A & $\times$ & $-\nu^2+K$ & $l \geq 1$ \\ \cline{3-9}
\rule[-3mm]{0mm}{8mm}&&${S}^{(n;lm)}_{ij}$ & odd and even & $\times$ & $\bigcirc$ & $-\nu^2+5K$ & $l \geq 1$  \\ \hline
\rule[-3mm]{0mm}{8mm}& tensor &$G^{(n;lm)}_{ij}$ & odd and even & $\times$ & $\times$ & $-\nu^2+2K$ & $l \geq 2$  \\
\hline \hline
\end{tabular}
\caption{
The properties of spherical and pseudo-spherical harmonics.
$K=\pm1$ denotes the spacial curvature.
The circle represents that the corresponding calculation can produce non-zero value.
On the other hand, the cross is denoted that the calculation always gives zero.
N/A means that the corresponding calculation is prohibited. 
}
\label{SandPS_table}
\end{center}
\end{table*}
 
\section{The formulae of harmonics in non-flat FLRW space}
\label{app_formulae}
In this section, the traces, covariant derivatives and norms of tensor spherical and pseudo-spherical harmonics 
in non-flat FLRW space are shown. 
If we refer spherical (pseudo-spherical) harmonics, namely, $K=1$ ($K=-1$) case, 
${\bf Y}$ is replaced $\ha{{\bf Y}}$ ($\ch{{\bf Y}}$).
Since both odd and even parity mode share the properties, 
we abbreviate the subscript of parity.

\subsection{gradients}
The gradients of the normalized scalar harmonics are given by
\begin{eqnarray}
\nabla_i Q^{(n;lm)} = \nu\, Q^{(n;lm)}_{i} \,,
\end{eqnarray}
where, $\nabla_i$ is a covariant derivative associated with the induced metric of non-flat FLRW space $g_{ij}$.
The symmetrized gradient of the normalized vector harmonics are given by
\begin{eqnarray}
\nabla_{(i} Q^{(n;lm)}_{j)} &=& - { \nu \over \sqrt{3} } Q^{(n;lm)}_{ij} +\sqrt{ {2 \over 3} (\nu^2 -3K) } P^{(n;lm)}_{ij} \,, \notag \\  \\
\nabla_{(i} S^{(n;lm)}_{j)} &=& \sqrt{ {\nu^2-3K \over 2} }S^{(n;lm)}_{ij} \,.
\end{eqnarray}

\subsection{traces}
The traces of the normalized harmonics are given by
\begin{eqnarray}
g^{ij} Q^{(n;lm)}_{ij} = { \sqrt{3} \over a^2} Q^{(n;lm)} \,,
\end{eqnarray}
and the others are all vanished.

\subsection{divergences}
The divergences of the normalized vector harmonics are vanished except $Q^{(n;lm)}_{i}$ :
\begin{eqnarray}
g^{ij}\nabla_i Q^{(n;lm)}_{j} &=& -{\nu \over a^2} Q^{(n;lm)} \,.
\end{eqnarray}
The non-trivial divergences of the normalized spherical harmonics are given by
\begin{eqnarray}
g^{jk}\nabla_k Q^{(n;lm)}_{ij} &=& {\nu \over \sqrt{3} \,a^2}  Q^{(n;lm)}_{ i} \,, \\ \notag  \\
g^{jk}\nabla_k P^{(n;lm)}_{ij} &=& -{1 \over a^2} \sqrt{  {2 (\nu^2-3K)  \over 3}  } Q^{(n;lm)}_{ i} \,,  \\ \notag \\ 
g^{jk}\nabla_k S^{(n;lm)}_{ij} &=&  -{1 \over a^2} \sqrt{  {\nu^2-3K  \over 2}  }  S^{(n;lm)}_{i} \,.
\end{eqnarray}

\subsection{Laplace-Beltrami operator}
The eigenvalues of the Laplace-Beltrami operator in non-flat FLRW space are listed.
Those of scalar type harmonics are given by
\begin{eqnarray}
\nabla^2 Q^{(n;lm)} &=& -{\nu^2 \over a^2 }Q^{(n;lm)} \,,   \\ \notag  \\
\nabla^2 Q^{(n;lm)}_i &=& -\left({\nu^2-2K \over a^2 }\right) Q^{(n;lm)}_i \, \\ \notag  \\
\nabla^2 Q^{(n;lm)}_{ij} &=& -{\nu^2 \over a^2 } Q^{(n;lm)}_{ij} \,, \\ \notag  \\
\nabla^2 P^{(n;lm)}_{ij} &=& -\left({\nu^2 -6K \over a^2 }\right) P^{(n;lm)}_{ij} \,, 
\end{eqnarray}
those of vector type are given by
\begin{eqnarray}
\nabla^2 S^{(n;lm)}_{i} &=& -\left({\nu^2 - K \over a^2 }\right) S^{(n;lm)}_{i} \,, \\ \notag  \\
\nabla^2 S^{(n;lm)}_{ij} &=& -\left( {\nu^2 -5K \over a^2 }\right) S^{(n;lm)}_{ij} \,,
\end{eqnarray}
and those of tensor type are given by
\begin{eqnarray}
\nabla^2 G^{(n;lm)}_{ij} &=& -\left( {\nu^2-2K \over a^2 }\right) G^{(n;lm)}_{ij} \,.
\end{eqnarray}

\section{Gauge fixing}
\label{sec_gauge_fixing}
The metric perturbations include both physical and gauge degrees of freedom.
Thus, we can perform further simplification by fixing gauge. 
Note that HL theory losses the general covariance because of Lifshitz scaling, 
i.e., the rotational transformation of time direction is prohibited. 
Thus, the infinitesimal coordinate transformation is expressed as follows : 
\begin{eqnarray}
t \to t + f(t)\,,~x^i \to x^i + \zeta^i (t,x^j) \,.
\end{eqnarray}
Consider the non-flat FLRW background :
\begin{eqnarray}
\bar{N}=1\,,~\bar{N}_i=0\,,~
\bar{g}_{ij} = 
\begin{cases}
a^2 \ha{\gamma}_{ij} & \mathrm{for}~K=1 \\
a^2 \ch{\gamma}_{ij} & \mathrm{for}~K=-1
\end{cases}\,.
\end{eqnarray}
Then, the infinitesimal transformations of the perturbed ADM quantities are given by
{
\begin{eqnarray}
\alpha^{(\mathrm{gauge})} &=& -\partial_t {f}\,,~ \\
\beta_i ^{(\mathrm{gauge})} &=& \partial_t {\zeta}_i -2 H \zeta_i \,, \\
h_{ij}^{(\mathrm{gauge})} &=& 2\nabla_{(i} \zeta_{j)} - 2 H f\, \bar{g}_{ij}  \,.
\end{eqnarray}
}
We define harmonic expansion of $\zeta^i$ as follows :  
\begin{eqnarray}
\zeta_i &=& \sum_{n,l,m} a^2 \Big[ \zeta^{(n;lm)}_{(Q)} Q_i^{(n;lm)} \notag \\
&&~~~~~~~+ \zeta^{(n;lm)}_{(S;o)} S_{(o)i}^{(n;lm)}  + \zeta^{(n;lm)}_{(S;e)} S_{(e)i}^{(n;lm)}   \Big]\,, 
\end{eqnarray}
Since $f$ does depend only on time, we do not have to expand by the harmonics. 
Then, we can explicitly describe the gauge transformations of perturbed ADM variables. 

We firstly consider the transformation of the scalar perturbation : 
\begin{eqnarray}
\alpha &\to& \alpha -\partial_t f \,,  \\ \notag \\
\beta^{(n;lm)}_{(Q)} &\to& \beta^{(n;lm)}_{(Q)}+ [\partial_t {-}2H]\zeta_{(Q)}^{(n;lm)} \,,  \\ \notag \\
h^{(n;lm)}_{(P)} &\to& h^{(n;lm)}_{(P)} +{2 \over a^2}\sqrt{ {2 (\nu^2 -3K) \over 3} }\zeta_{(Q)}^{(n;lm)}\,, 
\end{eqnarray}
and 
\begin{eqnarray}
&&\sum_{n,l,m}h_{(Q)}^{(n;lm)} Q^{(n;lm)}_{ij} \notag \\
&&\to \sum_{n,l,m} \left[ h^{(n;lm)}_{(Q)}-{2 \nu \over \sqrt{3} \, a^2}  \zeta_{(Q)}^{(n;lm)}  \right]Q^{(n;lm)}_{ij} \notag 
-2\sqrt{3} H f \bar{g}_{ij} \,, \notag \\
\label{trans_law_hQ}
\end{eqnarray}
Those of the vector perturbation are given by
\begin{eqnarray}
\beta^{(n;lm)}_{(S)} &\to& \beta^{(n;lm)}_{(S)} +[\partial_t {  -} 2H]\zeta_{(S)}^{(n;lm)}  \,,  \\ \notag \\
h^{(n;lm)}_{(S)} &\to& h^{(n;lm)}_{(S)} +\sqrt{2 (\nu^2 -3K) } \zeta_{(S)}^{(n;lm)}\,,  
\end{eqnarray}
Since both odd and  even parity modes obey the same transformation law, 
the parity subscripts are abbreviated.
Those of the tensor perturbation are given by:
\begin{eqnarray}
h^{(n;lm)}_{(G)} &\to& h^{(n;lm)}_{(G)} \,,
\end{eqnarray}
where, the parity subscripts are also abbreviated for the same reason as vector perturbation.
Note that the tensor perturbation is gauge invariant. 

Then, we consider the gauge fixing to simplify the procedure for perturbation.
In scalar perturbation, we find two types of quantities which can be manipulated, i.e., $f$ and $\zeta_{(Q)}$.
Obviously, $h_{(P)}$ which is the traceless part of scalar perturbation can be eliminated by choosing  
\begin{eqnarray}
\zeta^{(n;lm)}_{(Q)} = -{a^2 \over 2} \left[ {2 (\nu^2 -3K) \over 3} \right]^{-1/2} h^{(n;lm)}_{(P)} \,.
\end{eqnarray}
However, the trace part $h_{(Q)}$ is not. 
Because the transformation law (\ref{trans_law_hQ}) includes spacially homogeneous part
which is proportional to $f \bar{g}_{ij}$.
Since the terms which include $\dot{h}_{(Q)}$ join the quadratic action, 
the scalar degree of freedom is appeared in this theory unlike the case of GR.
Instead, we eliminate the lapse perturbation $\alpha$ by solving differential equation $\partial_t f =\alpha$.

In the vector perturbation, $\zeta_{(S;o)}$ and $\zeta_{(S;e)}$ can be manipulated, thus, 
we eliminate $h_{(S;o)}$ and $h_{(S;e)}$ by choosing
\begin{eqnarray}
\zeta^{(n;lm)}_{(S;o)} &=& -\left[2(\nu^2-3K) \right]^{-1/2} h_{(S;o)}^{(n;lm)}  \,,\\
\zeta^{(n;lm)}_{(S;e)} &=& -\left[2(\nu^2-3K) \right]^{-1/2} h_{(S;e)}^{(n;lm)}  \,.
\end{eqnarray}

To recap, we can eliminate the following perturbation modes by choosing gauge : 
\begin{eqnarray}
\alpha = h^{(n;lm)}_{(P)}= h^{(n;lm)}_{(S;o)}=h^{(n;lm)}_{(S;e)}= 0\,.
\end{eqnarray}

%%%%%%%%%%%%%%%%%%%%%%%%%%%%%%%%%%%%
%%%%%%%%%%%%%%%%%%%%%%%%%%%%%%%%%%%%%%

%======================================%
%<<<<<<<<<<<< BIBLIOGRAPHY >>>>>>>>>>>>%
%======================================%
%%%%%%%%%%%%%%%%%%%%%%%%%%%%%%%%%%%%%%%%%%%%%%%%%%
%%%%%%%%%%%%%%%%%%%%%%%%%%%%%%%%%%%%%%%%%%%%%%%%%%


\begin{thebibliography}{99}

 \bibitem{singularity_theorem}
 R.~Penrose, Phys. Rev. Lett. {\bf 14}, 57 (1965); 
 S.~W.~Hawking, Proc. R. Soc., A {\bf 300}, 187 (1967); 
 S.~W.~Hawking and R.~Penrose, Proc. R. Soc., A {\bf 314}, 529 (1970); 
 S.~W.~Hawking and G.~F.~R.~Ellis, {\it The Large Scale Structure of Space-Time} (Cambridge Univ., Cambridge, 1973).
 
 \bibitem{bouncing_cosmologies}
 See for example M.~Novello and S.~E.~P.~Bergliaffa, 
 ``bouncing cosmologies"
 Phys. Rep. {\bf 463}, 127 (2008), and references therein.
 
 \bibitem{ref_superstring}
 M.~B.~Green, J.~H.~Schwarz and E.~Witten, {\it Superstring Theory} (Cambridge University Press, Cambridge, 1987); 
 J.~Polchinski, {\it String Theory} (Cambridge University Press, Cambridge, 1998).
 
 \bibitem{ref_LQG}
 See for example, C.~Rovelli, {\it Quantum Gravity} (Cambridge University Press, Cambridge, 2004).
 
 \bibitem{ref_CDT}
 R.~Loll, {\it Discrete Lorentzian quantum gravity}, Nucl. Phys. B, Proc. Suppl. {\bf 94}, 96 (2001).
 
 \bibitem{ref_non-local}
 T.~Biswas, E.~Gerwick, T.~Koivisto and A.~Mazumdar, 
 ``Towards Singularity and Ghost-free Theories of Gravity"
 Phys. Rev. Lett. {\bf 108}, 031101 (2012);
 T. Biswas, A.~S.~Koshelev, A.~Mazumdar and S.~Y.~Vernov,
``Stable bounce and inflation in non-local higher derivative cosmology"
JCAP {\bf 08}, 024 (2012).
 
 \bibitem{HL_original}
 P.~Ho\v{r}ava, ``Quantum gravity at a Lifshitz point", Phys. Rev. D {\bf 79}, 084008 (2009).
 
  \bibitem{ref_LS}
 E.~M.~Lifshitz, ``On the Theory of Second-Order Phase Transitions I \& II", 
 Zh. Eksp. Teor. Fiz. {\bf 11} (1941) 255 \& 269.
  
 \bibitem{ref_renormalize_HL}
 A.~O.~Barvinsky, D.~Blas, M.~Herrero-Valea, S.~M.~Sibiryakov and  C.~F.~Steinwachs,
 ``Renormalization of Horava gravity", 
 Phys. Rev. D. {\bf93} . 064022 (2016) [arXiv:1512.02250[hep-th]].
 
 \bibitem{HL_BHTS}
 D.~Blas and S.~Sibiryakov, 
 ``Ho\v{r}ava gravity versus thermodynamics: The black hole case”, 
 Phys. Rev. D {\bf 84}, 124043 (2011) [arXiv:1110.2195 [hep-th]];
 E.~Barausse, T.~Jacobson and T.~P.~Sotiriou, 
 ``Black holes in Einstein-aether and Ho\v{r}ava-Lifshitz gravity”, 
 Phys. Rev. D {\bf 83}, 124043 (2011) [arXiv:1104.2889[gr-qc]]; 
 P.~Berglund, J.~Bhattacharyya and D.~Mattingly, 
 ``Mechanics of universal horizons", 
 Phys. Rev. D {\bf 85}, 124019 (2012) [arXiv:1202.4497[hep-th]]; 
 Y.~Misonoh and K. Maeda, 
 ``Black Holes and Thunderbolt Singularities with Lifshitz Scaling Terms", 
 Phys. Rev. D {\bf 92}, 084049 (2015) [arXiv:1509.01378[gr-qc]].
 
 \bibitem{matter_bounce_HL}
E.~Kiritsis and G.~Kofinas, 
``Horava-Lifshitz Cosmology”, 
Nucl. Phys. B{\bf 821}: 467-480 (2009)[arXiv:0904.1334[hep-th]];
R.~H.~Brandenberger, 
``Matter Bounce in Horava-Lifshitz Cosmology”, 
Phys. Rev. D{\bf 80}, 043516 (2009) [arXiv:0904.2835 [hep-th]].
  
  \bibitem{class_HL_cosmology}
  T.~Ha, Y.~Huang, Q.~Ma, K.~D.~Pechan,~T.~J.~Renner, Z.~Wu, G.~A.~Benesh and A.~Wang, 
  ``Classification of the FRW universe with a cosmological constant and a perfect fluid of the equation of state $p=w\rho$"
  Gen. Relativ. Grav. {\bf 44} 1433-1458 (2012) [arXiv:0905.0396[physisc.pop-ph]];
  A.~Wang and Y.~Wu, 
  ``Thermodynamics and classification of cosmological models in the Horava-Lifshitz theory of gravity"
  JCAP {\bf 0907} 012 (2009) [arXiv:0905.4117[hep-th]].
  
  
  \bibitem{previous_1}
K.~Maeda, Y.~Misonoh and T.~Kobayashi, 
``Oscillating Universe in Horava-Lifshitz Gravity"
Phys. Rev. {\bf D 82}, 064024 (2010).

\bibitem{previous_2}
Y.~Misonoh, K.~Maeda and T.~Kobayashi, 
``Oscillating Bianchi IX Universe in Horava-Lifshitz Gravity"
Phys. Rev. {\bf D 84}, 064030 (2011).

 \bibitem{HL_pert}
 R.~Cai, B.~Hu and H.~Zhang, 
 ``Dynamical Scalar Degree of Freedom in Horava-Lifshitz Gravity”, 
 Phys. Rev. D {\bf 80}, 041501 (2009) [arXiv:0905.0255 [hep-th]];
 K.~Yamamoto, T.~Kobayashi and G.~Nakamura, 
 ``Breaking the scale invariance of the primordial power spectrum in Horava-Lifshitz Cosmology”,
 Phys. Rev. D {\bf 80}, 063514 (2009) [arXiv:0907.1549 [astro- ph.CO]]; 
 Y.~Lu and Y.~Piao, 
 ``Scale Invariance from Modified Dispersion Relations”, 
 Int. J. Mod. Phys. D {\bf 19}, 1905 (2010) [arXiv:0907.3982 [hep-th]]; 
T.~Kobayashi, Y.~Urakawa and M.~Yamaguchi, 
``Large scale evolution of the curvature perturbation in Horava-Lifshitz cosmology", 
JCAP {\bf 0911}, 015 (2009) [arXiv:0908.1005 [astro-ph.CO]]; 
T.~Kobayashi, Y.~Urakawa and M.~Yamaguchi, 
``Cosmological perturbations in a healthy extension of Horava gravity", 
JCAP {\bf 1004}, 025 (2010) [arXiv:1002.3101 [hep-th]];
K.~Izumi, T.~Kobayashi and S.~Mukohyama, 
``Non-Gaussianity from Lifshitz Scalar", 
JCAP {\bf 1010}, 031 (2010) [arXiv:1008.1406 [hep-th]].

\bibitem{IR_stability}
C.~Charmousis, G.~Niz, A.~Padilla and P.~M.~Saffin, 
``Strong coupling in Horava gravity”, 
JHEP {\bf 0908}, 070 (2009) [arXiv:0905.2579 [hep-th]]; 
M.~Li and Y.~Pang, 
``A Trouble with Horava-Lifshitz Gravity", 
JHEP {\bf 0908} 015 (2009) [arXiv:0905.2751 [hep-th]];
D.~Blas, O.~Pujolas and S.~Sibiryakov, 
``On the Extra Mode and Inconsistency of Horava Gravity", 
JHEP {\bf 0910} 029 (2009) [arXiv:0906.3046 [hep-th]].
[hep-th];
C.~Bogdanos and E.~N.~Saridakis, 
``Perturbative instabilities in Horava gravity”, 
Class. Quant. Grav. {\bf 27}, 075005 (2010) [arXiv:0907.1636 [hep-th]]; 
S.~Mukohyama, 
``Horava-Lifshitz Cosmology: A Review", 
Class. Quant. Grav. {\bf 27} 223101 (2010) [arXiv: 1007.5199 [hep-th]].

\bibitem{healty_ex}
D.~Blas, O.~Pujolas and S.~Sibiryakov, 
``Consistent Extension of Ho\v{r}ava Gravity", 
Phys. Rev. Lett. {\bf 104}, 181302 (2010) [arXiv:0909.3525 [hep-th]].

\bibitem{dS_stability}
Y.~Q.~Huang, A.~Wang and Q.~Wu, 
``Stability of the de Sitter spacetime in Horava-Lifshitz theory",
Mod. Phys. Lett. {\bf A25} 2267 (2010) [arXiv:1003.2003[hep-th]];
A.~Wang and Q.~Wu,
``Stability of spin-0 graviton and strong coupling in Horava-Lifshitz theory of gravity",
Phys.~Rev.~D {\bf 83} 044025 (2011) [arXiv:1009.0268[hep-th]].

\bibitem{NF_pert_1}
A.~Wang and R.~Maartens, 
``Cosmological perturbations in Horava-Lifshitz theory without detailed balance”
Phys. Rev. D81, 024009 (2010) [arXiv:0907.1748 [hep-th]].

\bibitem{NF_pert_2}
X.~Gao,~Y.~Wang,~W.Xue and R.~H.~Brandenberger, 
``Fluctuations in a Horava-Lifshitz bouncing cosmology”
JCAP 1002, 020 (2010) [arXiv:0911.3196 [hep-th]]. 

  \bibitem{SVW}
  T.~P.~Sotiriou, M.~Visser and S.~Weinfurtner,
  ``Phenomenologically viable Lorentz-violating quantum gravity", 
  Phys. Rev. Lett.  {\bf 102}, 251601 (2009)
  [arXiv:0904.4464 [hep-th]];
    T.~P.~Sotiriou, M.~Visser and S.~Weinfurtner,
  ``Quantum gravity without Lorentz invariance", 
  JHEP {\bf 0910}, 033 (2009)
  [arXiv:0905.2798 [hep-th]].
  
 \bibitem{DM_integrate}
 S.~Mukohyama, 
 ``Dark matter as integration constant in Horava-Lifshitz gravity", 
 Phys. Rev. {\bf D 80}, 064005 (2009)[arXiv:0905.3563 [hep-th]].

\bibitem{Lifshitz_Khalatnikov}
E.~M.~Lifshitz and I.~M.~Khalatnikov, 
``Investigations in relativistic cosmology",
 Adv. Phys. {\bf 12}, 46 (1963).

\bibitem{Ynlm_ref}
V.~D.~Sandberg, 
``Tensor spherical harmonics on $S^2$ and $S^3$ as eigenvalues problems", 
J. Math. Phys. {\bf 19} (12) (1978).

\bibitem{SandPS}
K.~Tomita,  
``Tensor Spherical and Pseudo-Spherical Harmonics in Four-Dimensional Spaces", 
Prog. Theor. Phys. {\bf 68}, 1, 310 (1982).

\bibitem{cosmic_no-hair}
R.~M.~Wald, 
``Asymptotic behavior of homogeneous cosmological models in the presence of a positive cosmological constant", 
Phys. Rev. D 28, 2118 (1983);
I.~Moss and V.~Sahni, 
``Anisotropy in the chaotic inflationary universe", 
Phys. Lett. B 178, 159 (1986); 
J.~D.~Barrow, 
``Cosmic no-hair theorem and inflation", 
Phys. Lett. B 187, 12 (1987); 
L.~G.~Jensen and J.~A.~Stein-Schabes, 
``Is inflation natural?", 
Phys. Rev. D 35, 1146 (1987).

\bibitem{Horndeski_theory}
G.~W.~Horndeski, 
``Second-order scalar-tensor field equations in a four-dimensional space", 
Int. J. Theor. Phys. {\bf 10}, 363 (1974); 
T.~Kobayashi, M.~Yamaguchi and J.~Yokoyama, 
``Generalized G-inflation: Inflation with the most general second-order field equations", 
Prog. Theor. Phys. {\bf 126}, 511 (2011) [arXiv:1105.5723 [hep-th]].

\bibitem{no-go_Horndeski}
T.~Kobayashi,
``Generic instabilities of non-singular cosmologies in Horndeski theory: a no-go theorem",
Phys. Rev. D {\bf 94}, 043511 (2016)  [arXiv:1606.05831 [hep-th]];
A.~Ijjas and P.~J.~Steinhardt, 
``Classically stable non-singular cosmological bounces", 
Phys. Rev. Lett. {\it 117}, 121304 (2016) [arXiv:1606.08880 [gr-qc]].

\bibitem{bounce_EFT}
Y.~Cai, Y.~Wan, H-G.~Li, T.~Qiu and Y-S.~Piao, 
``The Effective Field Theory of nonsingular cosmology", 
arXiv:1610.03400 [gr-qc].

\bibitem{forthcoming_paper}
Y.~Misonoh, M.~Fukushima and S.~Miyashita, {\it in preparation}.

\bibitem{Bianchi_class}
See for example, M. P. Ryan and L. C. Shepley, {\it Homogeneous Relativistic Cosmologies} (Princeton University Press, 1975);
F.~Sato and H.~Kodama, {\it General Relativity} (Iwanami, 1992).

\bibitem{AandS}
M.~Abramowitz and I.~A.~Stegun,  
{\it Handbook of Mathematical Functions with Formulas, Graphs, and Mathematical Tables}
(Dover Publications, 1965).

\end{thebibliography}
\end{document}